\documentclass[twocolumn,superscriptaddress,amssymb,amsmath,nobibnotes,aps,prd,showpacs,nofootinbib]{revtex4-1}
\pdfoutput=1

\usepackage{graphicx,bm,color,psfrag,hyperref}
\usepackage{amsfonts}
\usepackage{lipsum}
\usepackage{caption}
\usepackage{subcaption}
\usepackage{newtxtext}
\usepackage{threeparttable,booktabs}

\newcommand{\be}{\begin{equation}}
\newcommand{\ee}{\end{equation}}

\begin{document}

\title{ Screening $\Lambda$ in a new modified gravity model }
\author{\"{O}zg\"{u}r Akarsu}
\email{akarsuo@itu.edu.tr}
\affiliation{Department of Physics, Istanbul Technical University, Maslak 34469 Istanbul, Turkey}

\author{John D. Barrow}
\email{J.D.Barrow@damtp.cam.ac.uk}
\affiliation{DAMTP, Centre for Mathematical Sciences, University of Cambridge, Wilberforce Road, Cambridge CB3 0WA, U.K.}

\author{Charles V. R. Board}
\email{cvrb2@damtp.cam.ac.uk}
\affiliation{DAMTP, Centre for Mathematical Sciences, University of Cambridge, Wilberforce Road, Cambridge CB3 0WA, U.K.}

\author{N. Merve Uzun}
\email{nebiye.uzun@boun.edu.tr}
\affiliation{Department of Physics, Bo\u{g}azi\c{c}i University, Bebek 34342 Istanbul, Turkey.}

\author{J. Alberto Vazquez}
\email{javazquez@icf.unam.mx}
\affiliation{Instituto de Ciencias F\'isicas, Universidad Nacional Aut\'onoma de M\'exico, Apdo. Postal 48-3, 62251 Cuernavaca, Morelos, M\'exico}
\begin{abstract}
We study a new model of Energy-Momentum Squared Gravity (EMSG), called Energy-Momentum Log Gravity (EMLG), constructed by the addition of the term $f(T_{\mu\nu}T^{\mu\nu})=\alpha \ln(\lambda\,T_{\mu\nu}T^{\mu\nu})$, envisaged as a correction, to the Einstein-Hilbert action with cosmological constant $\Lambda$. The choice of this modification is made as a specific way of including new terms in the right-hand side of the Einstein field equations, resulting in constant effective inertial mass density and, importantly, leading to an explicit exact solution of the matter energy density in terms of redshift. We look for viable cosmologies, in particular, an extension of the standard $\Lambda$CDM model. EMLG provides an effective dynamical dark energy passing below zero at large redshifts, accommodating a mechanism for screening $\Lambda$ in this region, in line with suggestions for alleviating some of the tensions that arise between observational data sets within the standard $\Lambda$CDM model. We present a detailed theoretical investigation of the model and then constrain the free parameter $\alpha'$, a normalisation of $\alpha$, using the latest observational data. The data does not rule out the $\Lambda$CDM limit of our model ($\alpha'= 0$), but prefers slightly negative values of the EMLG model parameter ($\alpha'= -0.032\pm 0.043$), which leads to the screening of $\Lambda$. We also discuss how EMLG relaxes the persistent tension that appears in the measurements of $H_0$ within the standard $\Lambda$CDM model.

\end{abstract}
\date{\today}


\maketitle

\section{Introduction}
\label{Sec:Intro}

The standard Lambda cold dark matter ($\Lambda $CDM) model is the most successful and economical cosmological model that accounts for the dynamics and the large-scale structure of the observable universe. Furthermore, it is in good agreement with the most of the currently available data \cite{Komatsu:2010fb,Planck2015,Aghanim:2018eyx}. Nevertheless, it suffers from profound theoretical issues relating to the cosmological constant $\Lambda $ \cite{Weinberg:1988cp,Peebles:2002gy,Padmanabhan:2002ji} and, on the observational side, from tensions of various degrees of significance between some existing data sets \cite{JAVazquez, tension01,tension02,Tamayo:2019gqj,Hee:2016ce,tension03,Aubourg:2014yra,Zhao:2017cud,Bullock:2017xww,Freedman:2017yms}. Firstly, the value of $H_{0}$ measured from the cosmic microwave background (CMB) data by the Planck Collaboration \cite{Planck2015} in the basic $\Lambda $CDM model is 3.4 $\sigma $ lower than the model-independent local value reported from supernovae by Riess et al. \cite{Riess:2016jrr}; secondly, the Lyman-$\alpha $ forest measurements of the baryon acoustic oscillations (BAO) by the Baryon Oscillation Spectroscopic Survey (BOSS) prefer a smaller value of the pressureless matter density parameter than is preferred by the CMB data within $\Lambda $CDM \cite{Delubac:2014aqe}. Such tensions are of great importance since detection of even small deviations from $\Lambda $CDM could imply profound modifications to the fundamental theories underpinning this model. For instance, the BOSS collaboration reported a clear detection of dark energy (DE) in \cite{Aubourg:2014yra}, consistent with positive $\Lambda $ for $z<1$, but with a preference for a DE yielding negative energy density values for $z>1.6$. They then argued that the Lyman-$\alpha $ data from $z\sim 2.3$ can be accommodated by a non-monotonic evolution of $H(z)$, and thus of $\rho _{\mathrm{tot}}(z)$ within general relativity (GR), which is difficult to realize in any model with non-negative DE density. However, a \textit{physical} DE with negative energy density would be physically problematic, which suggests that DE might instead be an \textit{effective source} arising from a modified theory of gravity (see \cite{Copeland:2006wr,Caldwell:2009ix,Clifton:2011jh,DeFelice:2010aj,Capozziello:2011et,Nojiri:2017ncd,Nojiri:2010wj} for reviews on DE and modified theories of gravity). In line with this, \cite{Sahni:2014dee} argues that the Lyman-$\alpha$ data can be addressed using a physically motivated modified gravity model that alters the Friedmann equation for $H(z)$ itself, and that a further tension, also relevant to the Lyman-$\alpha $ data, can be alleviated in models in which $\Lambda$ is dynamically screened, implying an effective DE passing below zero and concurrently exhibiting a pole in its equation of state (EOS) at large redshifts. The possible modifications to the $H(z)$ of $\Lambda $CDM can be represented by $3H^{2}(z)=\rho _{\mathrm{m,0}}(1+z)^{3}[1-u(z)]+\Lambda-v(z),$ involving functions $u(z)$ and $v(z)$ that represent two principal modifications. Interpreting all the terms other than $\rho _{\mathrm{m,0}}(1+z)^{3}$ as arising from DE, i.e. writing $3H^{2}(z)=\rho _{\mathrm{m,0}}(1+z)^{3}+\rho _{\mathrm{DE}}$, would lead to an effective DE of the form $\rho _{\mathrm{DE}}=\Lambda -\rho _{\mathrm{m,0}}u(z)(1+z)^{3}-v(z)$. Accordingly, $u(z)>0$ and $v(z)>0$ would drive $\rho _{\mathrm{DE}}$ towards negative values, and so $\Lambda $ could be screened and $\rho _{\mathrm{DE}}<0$ when we have $\rho _{\mathrm{m,0}}u(z)(1+z)^{3}+v(z)>\Lambda $. Dynamical $u(z)$ and $v(z)$ functions are familiar from scalar-tensor theories, in which $u(z)$ stands for a varying effective gravitational coupling strength in the Jordan frame (or non-conservation, say, of the pressureless matter in the Einstein frame \cite{Faraoni:1998qx}), while $v(z) $ stands for the new terms due to the scalar field associated with varying gravitational `constant', $G$. In such models, when the effective gravitational coupling strength gets weaker with increasing redshift, $\rho_{\mathrm{DE}}$ (as defined above) becomes negative at large redshifts \cite{Faraoni:1998qx,Boisseau:2000pr,Sahni:2006pa,Akarsu:2019pvi}. A range of other examples of $\rho _{\mathrm{DE}}$ crossing below zero exist, including theories in which $\Lambda $ relaxes from a large initial value via an adjustment mechanism \cite{Dolgov:1982qqA,Dolgov:1982qqB,Bauer:2010wj}, cosmological models based on Gauss-Bonnet gravity \cite{Zhou:2009cy}, braneworld models \cite{Sahni:2002dx,Brax:2003fv}, loop quantum cosmology \cite{Ashtekar:2006wn,Ashtekar:2011ni}, and higher dimensional cosmologies that accommodate dynamical reduction of the internal space \cite{Chodos:1979vk,Dereli:1982ar,Akarsu:2012am,Akarsu:2012vv,Russo:2018akp}. In this paper, as a new example of such zero-crossing models, we study a particular theory of modified gravity: Energy-Momentum Squared Gravity (EMSG) \cite{Katirci:2014sti,Roshan:2016mbt,Akarsu:2017ohj,Board:2017ign,Akarsu:2018zxl,Nari:2018aqs,Akarsu:2018aro}, which generalizes the form of the matter Lagrangian in a non-linear way and ensures that both $u(z)$ and $v(z)$ are dynamical. We will make a specific choice of model within the theory, in order to establish whether it is a good candidate for such behaviour.

From the Einstein-Hilbert action of GR, it is possible to consider a generalisation involving non-linear matter terms, by adding some analytic function of a new scalar $T^{2}=T_{\mu \nu }T^{\mu \nu }$ formed from the energy-momentum tensor (EMT), $T_{\mu \nu }$, of the matter stresses \cite{Katirci:2014sti}. Such generalizations of GR result in new contributions by the usual material stresses to the right-hand side of the generalised Einstein field equations, $v(z)$, and lead in general to non-conservation of the material stresses, $u(z)$, without the need to invoke new forms of matter (for other similar types of theories, \cite{Harko:2010mv,Harko:2011kv}). A particular example of EMSG is when $f(T^{2})=\alpha T^{2}$, which has been studied in various contexts in \cite{Roshan:2016mbt,Board:2017ign,Akarsu:2018zxl,Nari:2018aqs}. EMSG of this form in the presence of dust leads to $u(z)=0$ and $v(z)=-\alpha \rho _{\mathrm{m}}^{2}=-\alpha \rho _{\mathrm{m,0}}^{2}(1+z)^{6}>0$ for $\alpha <0$, as in loop quantum cosmology \cite{Ashtekar:2006wn,Ashtekar:2011ni}, which would lead to negative DE in the past, whilst the case $\alpha >0$ corresponds to the braneworld scenarios \cite{Brax:2003fv}. However, if the quadratic energy density term is large enough to be effective today, then it would be the dominant term after just a few redshift units from today ($z=0$) and hence spoil the successful description of the early universe.

A generalisation of the above model with $f(T^{2})=\alpha T^{2}$, is Energy-Momentum Powered Gravity (EMPG), where $f(T^{2})=\alpha (T^{2})^{\eta }$, as studied in \cite{Akarsu:2017ohj,Board:2017ign}. This modification becomes effective at high energy densities, as in the early universe \cite{Board:2017ign,Akarsu:2018zxl}, for the cases with $\eta >1/2$, and at low energy densities, as in the late universe, when $\eta <1/2$ \cite{Akarsu:2017ohj}. For instance, $\eta =0$ leads mathematically to exactly the same background dynamics as $\Lambda $CDM and $\eta \simeq 0$ to a $w$CDM-type cosmological model, despite the only physical source in the model being dust \cite{Akarsu:2017ohj}. A recent study constraining the model from the low-redshift cosmological data can be found in \cite{Faria:2019ejh} and a dynamical systems analysis  in \cite{Bahamonde:2019urw}. EMPG results in both $u(z)$ and $v(z)$ arising dynamically and could be investigated for producing effective DE passage below zero at large redshifts. Nevertheless, it is generally not possible to obtain explicit exact solutions for $\rho _{\mathrm{m}}(z),$ and hence of $\rho _{\mathrm{de}}(z)$, which renders EMPG inconvenient for the present study \cite{Akarsu:2017ohj,Board:2017ign}. The particular case $\eta =1/2$, dubbed, `Scale Independent EMSG', is one of the exceptions, along with the case $\eta =1$ (EMSG with $f(T^{2})=\alpha T^{2}$), which provides explicit exact solutions for $H(z)$ required for a detailed observational test. In this model, the new terms in the field equations enter with the same power as the usual terms in GR, yet the standard energy is not conserved, and this leads to $u(z)=(1+z)^{3\alpha }-1$ and $v(z)=\rho_{\mathrm{m,0}}(1+z)^{3+3\alpha }$, which could provide the desired features in the $\alpha <0$ case. Nevertheless, this model is studied in detail in \cite{Akarsu:2018aro} (though in somewhat different context) and $\alpha $ is well constrained observationally to be so close to zero that Scale Independent EMSG is unable to resolve the issues noted above.

In what follows we consider a new type of EMSG, called Energy-Momentum Log Gravity, EMLG, constructed by the choice of $f(T_{\mu\nu}T^{\mu\nu})=\alpha \ln(\lambda\,T_{\mu\nu}T^{\mu\nu})$, where $\lambda>0$ and $\alpha$ are real constants, to the Einstein-Hilbert action with cosmological constant $\Lambda $. \footnote{A related logarithmic modification is considered in the context of $f(R,T)$ gravity \cite{Harko:2011kv} (where $T=g^{\mu\nu}T_{\mu\nu}$) in a recent paper \cite{Elizalde:2019ote} after our work. They extend the Starobinsky action \cite{Starobinsky:1980teA,Starobinsky:1980teB} by including the logarithmic trace of the energy-momentum tensor, $f(T)\propto \ln(T)$, and study the cosmological dynamics.} This form, which determines $u(z)$ and $v(z)$ in a specific way depending on $\alpha $, has appealing features. It gives rise to new contributions that appear similar to those of a perfect fluid with constant equation of state parameter on the right-hand side of the Friedmann equations, reminiscent of a source with constant inertial mass density, and furthermore it allows us to obtain an explicit exact solution of the pressureless matter energy density in terms of redshift, so that we can conduct an exact theoretical investigation of the model using the observational data without further simplifications. We look for observationally viable cosmologies, in particular, for an extension of the standard $\Lambda $CDM model. We find that the observational data does not exclude the $\Lambda $CDM limit of our model but slightly prefers $u(z)>0$ (related to the non-conservation of pressureless matter) and $v(z)<0$ (related to the new terms of the pressureless matter in the field equations), where $u(z)>0$ arises with the appropriate sign to produce an effective dynamical DE passing below zero (a screening of $\Lambda $) at high redshifts, as desired to address the tension with the Lyman-$\alpha $ measurements within the standard $\Lambda $CDM model. We also discuss the fact that the EMLG model relaxes, at some level, the persistent tension that appears between different measurements of $H_{0}$ within the standard $\Lambda $CDM model.

\section{Energy-Momentum Log Gravity}
\label{sec:EMLG}
We begin with the action constructed by the addition of the term $f(T_{\mu\nu}T^{\mu\nu})$ to the Einstein-Hilbert (EH) action with a cosmological constant, $\Lambda$, as follows
\begin{align}
S=\int \left[\frac{1}{2\kappa}(R-2 \Lambda)+f(T_{\mu\nu}T^{\mu\nu})+\mathcal{L}_{\rm m}\right]\sqrt{-g}\,{\rm d}^4x,
\label{action}
\end{align}
where $\kappa$ is Newton's constant scaled by a factor of $8\pi$ (and we henceforth set $\kappa=1$), $R$ is the Ricci scalar, $g$ is the determinant of the metric $g_{\mu\nu}$, $\mathcal{L}_{\rm m}$ is the Lagrangian density corresponding to the matter source described by the energy-momentum tensor $T_{\mu\nu}$, and we have used units such that $c=1$. We retain the cosmological constant, $\Lambda$, in the model since according to Lovelock's theorem it arises as a constant of nature. \footnote{Lovelock's theorem \cite{Lovelock:1971yv,Lovelock:1972vz} states that the only possible second-order Euler-Lagrange expression obtainable in a four-dimensional space from a scalar density of the form $\mathcal{L}= \mathcal{L}(g_{\mu\nu})$ is $E_{\mu\nu}=\sqrt{-g}\left(\lambda_1 G_{\mu\nu}+\lambda_2 g_{\mu\nu}\right)$, where $\lambda_1$ and $\lambda_2$ are constants, leading to Newton's gravitational constant $G\equiv\kappa/8\pi$ and cosmological constant $\Lambda$ in Einstein's field equations $G_{\mu\nu}+\Lambda g_{\mu\nu}=\kappa T_{\mu\nu}$ (see \cite{Bull:2015stt,Clifton:2011jh,Straumann} for further reading).}

We take the variation of the action with respect to the inverse metric $g^{\mu\nu}$ as

\begin{align}
  \delta S=\int\, {\rm d}^4 x \sqrt{-g} \bigg[&\frac{1}{2}\delta R+\frac{\partial f}{\partial(T_{\mu\nu}T^{\mu\nu})}\frac{\delta(T_{\sigma\epsilon}T^{\sigma\epsilon})}{\delta g^{\mu\nu}}\delta g^{\mu\nu} \nonumber \\
  &-\frac{1}{2}g_{\mu\nu}\left(\frac{R}{2}-\Lambda+f(T_{\sigma\epsilon}T^{\sigma\epsilon})\right)\delta g^{\mu\nu} \nonumber \\
  &+\frac{1}{\sqrt{-g}}\frac{\delta(\sqrt{-g}\mathcal{L}_{\rm m})}{\delta g^{\mu\nu}}\delta g^{\mu\nu}\bigg],
\end{align}

  and, as usual, we define the EMT in terms of the matter Lagrangian $\mathcal{L}_{\rm m}$ as follows
  \begin{equation}
  \label{tmunudef}
 T_{\mu\nu}=-\frac{2}{\sqrt{-g}}\frac{\delta(\sqrt{-g}\mathcal{L}_{\rm m})}{\delta g^{\mu\nu}}=g_{\mu\nu}\mathcal{L}_{\rm m}-2\frac{\partial \mathcal{L}_{\rm m}}{\partial g^{\mu\nu}}.
 \end{equation}

Accordingly, the modified Einstein field equations read
\begin{equation}
G_{\mu\nu}+\Lambda g_{\mu\nu} = T_{\mu\nu}+fg_{\mu\nu}-2\frac{\partial f}{\partial(T_{\mu\nu}T^{\mu\nu})}\theta_{\mu\nu},
\label{genfieldeq}
\end{equation}
where $G_{\mu\nu}=R_{\mu\nu}-\frac{1}{2}Rg_{\mu\nu}$ is the Einstein tensor and $\theta_{\mu\nu}$ is a new tensor defined as
\begin{equation}
\begin{aligned}
\theta_{\mu\nu}&= T^{\sigma\epsilon}\frac{\delta T_{\sigma\epsilon}}{\delta g^{\mu\nu}}+T_{\sigma\epsilon}\frac{\delta T^{\sigma\epsilon}}{\delta g^{\mu\nu}}  \\
&=-2\mathcal{L}_{\rm m}\left(T_{\mu\nu}-\frac{1}{2}g_{\mu\nu}\mathcal{T}\right)-\mathcal{T}T_{\mu\nu}\\
&\quad\quad+2T_{\mu}^{\gamma}T_{\nu\gamma}-4T^{\sigma\epsilon}\frac{\partial^2 \mathcal{L}_{\rm m}}{\partial g^{\mu\nu} \partial g^{\sigma\epsilon}}
\label{theta}
\end{aligned}
\end{equation}
with $\mathcal{T}$ being the trace of the EMT, $T_{\mu\nu}$.  We note that the EMT given in \eqref{tmunudef} does not include the second variation of $\mathcal{L}_{\rm m}$, and hence the last term of \eqref{theta} vanishes. As the definition of the matter Lagrangian that gives rise to the perfect-fluid EMT is not unique, one could choose either $\mathcal{L}_{\rm m}=p$ or $\mathcal{L}_{\rm m}=-\rho$, which result in the same EMT. In the present study, we consider $\mathcal{L}_{\rm m}=p$.

We proceed with a specific form of the model,
\begin{equation}
\label{eqn:lnassumption}
f(T_{\mu\nu}T^{\mu\nu})=\alpha \ln(\lambda\, T_{\mu\nu}T^{\mu\nu}),
\end{equation}
where $\lambda$ has the dimension inverse energy density squared so that $\lambda\, T_{\mu\nu}T^{\mu\nu}$ is dimensionless. This choice comes with some particular advantageous features. In the cosmological application of the model, this is the only functional choice of $f(T_{\mu\nu}T^{\mu\nu})$ that gives rise to new contributions of a perfect fluid on the right hand side of the Einstein field equations yielding constant effective inertial mass density (See Section \ref{sec:cimd} for details). Also, it has an explicit exact solution, including the form of $\rho(z)$ which is important for analytical investigations. This contrasts with many EMSG-type models, in which this is usually not possible due to the non-linear coupling of the matter sources to gravity. For instance, in \cite{Akarsu:2017ohj} cosmic acceleration in a dust only EMPG model was investigated, where the exact solution of $z(\rho_{\rm m})$ was obtained, but the corresponding explicit solution of $\rho_{\rm m}(z)$ could usually only be obtained through an approximation procedure, except for a few particular cases (\cite{Board:2017ign,Akarsu:2018aro}).

Consequently, the action we use is 
\begin{equation}
S=\int \left[\frac{1}{2}(R-2\Lambda)+\alpha \ln(\lambda \,T_{\mu\nu}T^{\mu\nu})+\mathcal{L}_{\rm m}\right]\sqrt{-g}\,{\rm d}^4x,
\label{eq:action}
\end{equation}
where $\alpha$ is a constant that determines the gravitational coupling strength of the EMLG modification of GR. Accordingly, the modified Einstein field equations \eqref{genfieldeq} for this action now read,
\begin{equation}
G_{\mu\nu}+\Lambda g_{\mu\nu}= T_{\mu\nu}+ \alpha g_{\mu\nu} \ln(\lambda\, T_{\sigma\epsilon}T^{\sigma\epsilon})-2 \alpha \frac{\theta_{\mu\nu}}{(T_{\sigma\epsilon}T^{\sigma\epsilon})}.
\label{fieldeq}
\end{equation}
From \eqref{fieldeq}, the covariant divergence of the EMT becomes
\begin{equation}
\label{nonconservedenergy}
\nabla^{\mu}T_{\mu\nu}=-\alpha g_{\mu\nu}\nabla^{\mu}\ln(\lambda\, T_{\sigma\epsilon}T^{\sigma\epsilon})
+2\alpha\nabla^{\mu}\left(\frac{\theta_{\mu\nu}}{T_{\sigma\epsilon}T^{\sigma\epsilon}}\right).
\end{equation}
We note that, unless $\alpha=0$, the right-hand side of this equation does not vanish in general, and thus the EMT is not conserved, i.e. $\nabla^{\mu}T_{\mu\nu}=0$ is not satisfied.

\section{Cosmology in EMLG}
In this paper, we investigate the cosmological behaviour of this gravitational model. We proceed by considering the spatially maximally symmetric spacetime metric, given by the Friedmann metric,
\begin{equation}
\label{RW}
{\rm d}s^2=-{\rm d}t^2+a^2\,\left[\frac{{\rm d}r^2}{1-kr^2}+r^2({\rm d}\theta^2+\sin^2\theta {\rm d}\phi^2)\right],
\end{equation}  
where the spatial curvature parameter $k$ takes values in $\{-1,\,0,\,1\}$ corresponding to open, flat and closed 3-spaces respectively, and the scale factor $a=a(t)$ is a function of cosmic time $t$ only. For cosmological matter sources describing the physical component of the universe, we consider the perfect fluid form of the EMT given by
\begin{equation}
\label{em}
T_{\mu\nu}=(\rho+p)u_{\mu}u_{\nu}+p g_{\mu\nu},
\end{equation} 
where $\rho>0$ is the energy density and $p$ is the thermodynamic pressure satisfying the barotropic equation of state (EoS) as
\begin{equation}
\label{eos}
\frac{p}{\rho}=w={\rm constant},
\end{equation}
and $u_{\mu}$ is the four-velocity satisfying the conditions $u_{\mu}u^{\mu}=-1$, and $\nabla_{\nu}u^{\mu}u_{\mu}=0$.

Using \eqref{em} and \eqref{eos}, we calculate $\theta_{\mu\nu}$ defined in \eqref{theta} and the self-contraction of the EMT for the perfect fluid with barotropic EoS \eqref{eos} as follows
\begin{align}
\label{thetafrw}
\theta_{\mu\nu}&=-\rho^2(3w+1)(w+1)u_{\mu}u_{\nu},\\
 T_{\mu\nu}T^{\mu\nu}&=\rho^2(3w^2+1).
\label{trace}
\end{align}
Next, using \eqref{thetafrw} and \eqref{trace} along with the metric \eqref{RW} in the modified Einstein field equations \eqref{fieldeq} we obtain the following pair of linearly independent modified Friedmann equations, for a single fluid cosmology,
\begin{align}
&3H^{2}+\frac{3k}{a^2}=\rho+\Lambda+\alpha' \rho_{0}+ \alpha' \rho_{0} \frac{2}{\gamma} \ln\left(\rho/\rho_{0}\right) , \label{eq:rhoprime}\\
&-2\dot{H}-3H^{2}-\frac{k}{a^2}= w \rho-\Lambda\nonumber\\
&\quad\quad\quad\quad\quad\quad\quad - \alpha' \rho_{0} \frac{2}{\gamma}\ln\left[\sqrt{3w^{2}+1}\;\left(\rho/\rho_{0}\right)\right],\label{eq:presprime}
\end{align}
where we set $\lambda=\rho_0^{-2}$ without loss of generality.\footnote{Defining $\lambda=\eta \rho_0^{-2}$, where $\eta>0$ is a coefficient, we can write $ \ln(\lambda \,T_{\mu\nu}T^{\mu\nu})= \ln(\eta)+\ln(T_{\mu\nu}T^{\mu\nu}/\rho_0^2)$. The term $\alpha \ln(\eta)$ then acts like a cosmological constant, and so simply rescales $\Lambda$ in the action \eqref{eq:action} and field equations \eqref{fieldeq}. Additionally, $\lambda$ has no contribution to the continuity equation \eqref{nonconservedenergy} since $\nabla^{\mu}\ln(\lambda\, T_{\sigma\epsilon}T^{\sigma\epsilon})=\nabla^{\mu}\ln( T_{\sigma\epsilon}T^{\sigma\epsilon})$. Therefore, choosing a particular value for $\eta$, i.e. $\eta=1$ as we have done, does not lead to any loss of generality as our model already includes $\Lambda$ in the action.} Here $H=\dot{a}/a$ is the Hubble parameter and the subscript $_0$ denotes the present-day values of the parameters. $\gamma=\gamma(w)$ is a parameter defined by
\begin{equation}
\gamma=\ln(3w^{2}+1)- 2 \frac{(3w+1)(w+1)}{(3w^{2}+1)},
\end{equation}
which is negative for $-0.27<w<2.52$ and positive otherwise. We also define the dimensionless constant
\begin{equation}
\alpha'= -\alpha\, \gamma \rho_{0}^{-1}.       
\label{eq:alphaprime}
\end{equation}
Note that in the action \eqref{eq:action}, the  terms $\alpha\ln(\lambda \,T_{\mu\nu}T^{\mu\nu})$ and $\mathcal{L}_{\rm m}$ are both related to the material content of the universe and that the EMT included in the modification term $\alpha\ln(\lambda \,T_{\mu\nu}T^{\mu\nu})$ is the same as the one obtained from the variation of $\mathcal{L}_{\rm m}$, so the model contains only a single matter source. However, the terms arising due to the EMLG modification couple to gravity with a different strength, $\alpha'$, to the normalized gravitational coupling strength (i.e. $\kappa=1$) of the standard GR terms. Furthermore, we note that $\alpha'$ is a function not only of the true constant of the EMLG modification, $\alpha$, but also the current energy density, $\rho_0$, and the EoS parameter, $w$, describing the type of the matter source, so $\alpha'=\alpha'(\alpha,\rho_{0},w)$. The latter two dependencies imply a violation of the equivalence principle, which means our modification must obey constraints from solar system tests of this principle. It would also have implications in fundamental physics. For example, the violation of equivalence principle is intimately connected with some of the basic aspects of the unification of gravity with particle physics such as string theories \cite{Uzan:2010pm} and theories of varying constants \cite{SBM1, SBM2, BM}. The consequences of this property of the model are beyond the scope of the current study, which focuses on the dynamics of a mono-fluid universe, where the only material source is dust (pressureless fluid) with the purpose of modifying $\Lambda$CDM by considering the new terms arising from EMLG as a correction.

The corresponding local energy-momentum conservation equation \eqref{nonconservedenergy} is
\begin{equation}
\label{noncons}
\dot{\rho}+3H(1+w)\rho\left[\frac{\gamma \rho(3w^2+1)-2 \alpha' \rho_{0}(3w+1)}{\gamma \rho(3w^2+1)+2 \alpha' \rho_{0}(3w^2+1)} \right]=0. 
\end{equation}
The expression in square brackets is the modification arising from EMLG and is equal to unity in the case $\alpha'=0$, corresponding to GR. We can see that the covariant energy-momentum conservation $\nabla^{\mu} T_{\mu\nu}=0$, which in GR would lead to $\rho\propto a^{-3(1+w)}$, does not hold for any $w\neq-1$ when $\alpha'\neq 0$, whilst the case $w=-1$, corresponding to conventional dark energy, i.e., vacuum energy, is unmodified by EMLG.

\subsection{Constant effective inertial mass density}
\label{sec:cimd}

It is worth noting here that, for a perfect fluid with barotropic equation of state, both $\theta_{\mu\nu}$ and $T_{\mu\nu}T^{\mu\nu}$ are proportional to $\rho^2$ and therefore the last term in \eqref{fieldeq} is independent of the energy density scale, instead depending only on the four-velocity of the fluid and type of the fluid (i.e., the EoS of the matter source). Furthermore, for usual cosmological applications, when a comoving (i.e. $u_{\mu}u^{\mu}=-1$ and $\nabla_{\nu}u^{\mu}u_{\mu}=0$) fluid with a constant EoS parameter $w$ is considered, this term becomes a constant determined by the model parameter $\alpha$ and the equation of state under consideration. On the other hand, the second term on the right-hand side of \eqref{fieldeq} will always contribute equally but with opposite signs to the time and space components of the equation in Lorentzian spacetimes, that is to the energy density and pressure equations arising from the metric given in \eqref{RW}, and therefore the addition of these equations results in the modifications from the second term on the right-hand side of \eqref{fieldeq} cancelling each other. Consequently, this produces a characteristic feature of the model: if we define the new terms that arise due to the EMLG modification in the energy density equation \eqref{eq:rhoprime} as an effective energy density 
\begin{equation}
\rho'=\alpha' \rho_{0}+ \alpha' \rho_{0} \frac{2}{\gamma} \ln\left(\rho/\rho_{0}\right) 
\end{equation}
and those in the pressure equation \eqref{eq:presprime} as an effective pressure  
\begin{equation}
p'=-\alpha' \rho_{0} \frac{1}{\gamma} \ln(3w^{2}+1)-\alpha' \rho_{0}\frac{2}{\gamma} \ln\left(\rho/\rho_{0}\right), 
\end{equation}
then the \textit{effective inertial mass density} defined as $\rho'+p'$ is always constant; specifically, 
\begin{equation}
\rho'+p'=\alpha' \rho_{0} [1-\gamma^{-1} \ln(3w^{2}+1)], 
\end{equation}
for $p/\rho=w={\rm constant}$. This feature of the model leads to $\rho'=\alpha' \rho_{0} [1-\gamma^{-1} \ln(3w^{2}+1)]-p'$ meaning that $\rho'$ changes sign when $p'=\alpha' \rho_{0} [1-\gamma^{-1} \ln(3w^{2}+1)]$, showing our model's relevance to the studies \cite{Sahni:2014dee,Aubourg:2014yra} suggesting that a DE model achieving negative energy density values for redshifts larger than a certain value (e.g., $z\gtrsim 2$ as suggested by \cite{Sahni:2014dee,Tamayo:2019gqj,Aubourg:2014yra}) might improve the fit to observational data. It might be mentioned that the sign change of $\rho'$ does not signal any pathologies since it is an \textit{effective} energy density, not the physical energy density.  For example, in the case of dust, $w=0$, we have
\begin{equation}
\rho'=\alpha'\rho_{\rm m,0}-p',
\end{equation}
and accordingly $\rho'<0$ when $p'>\alpha'\rho_{\rm m,0}$.

\subsection{Preliminary constraints on $\alpha$}
\label{sec:preliminary}

We now determine some preliminary constraints on $\alpha$ by considering separately two standard cosmological matter sources: radiation and dust. We begin by writing \eqref{noncons} in terms of $\alpha$:
\begin{equation}   \label{cont1}
\dot{\rho}=-3(1+w) H \rho\left[\frac{ \rho(3w^2+1)+2 \alpha(3w+1)}{ \rho(3w^2+1)-2 \alpha(3w^2+1)} \right]. 
\end{equation}
A viable cosmological model should satisfy $H>0$,\;$\dot{H}<0$,\;$\rho>0$ and $\dot{\rho}<0$. Here $H>0$ and $\dot{H}<0$ together lead to an expanding universe in line with observations. $\dot{\rho}<0$ means that the energy density is decreasing with time, and therefore $H>0$ and $\dot{\rho}<0$ together guarantee that the density is larger at early times and decreases as the universe expands. As seen from \eqref{cont1}, taking $H>0$, \;$\dot{\rho}<0$ implies
\begin{equation}  \label{decr.dens.}
(1+w) \rho\left[\frac{ \rho(3w^2+1)+2 \alpha(3w+1)}{ \rho(3w^2+1)-2 \alpha(3w^2+1)} \right]>0.
\end{equation}
Substituting $w=1/3$ into \eqref{decr.dens.}, we obtain the interval
\begin{equation}     \label{int-rad}
-\frac{\rho_{\rm r}}{3}<\alpha<\frac{\rho_{\rm r}}{2}
\end{equation}
over which it is guaranteed that the energy density of radiation, $\rho_{\rm r}$, increases as we go to earlier times. Next, we also substitute $w=0$ into \eqref{decr.dens.} and obtain the interval
\begin{equation}
\label{cond:dust}
-\frac{\rho_{\rm m}}{2}<\alpha<\frac{\rho_{\rm m}}{2}
\end{equation}
over which it is guaranteed that $\rho_{\rm m}$ (energy density of dust) decreases as the universe expands. From \eqref{eq:rhoprime} and \eqref{eq:presprime}, one can see that the energy density corresponding to the spatial curvature evolves as $\rho_{k}=\frac{3k}{a^2}$. We note that this is equivalent to a matter source with an EoS parameter $w=-1/3$ via $\nabla^{\mu} T_{\mu\nu}=0$ in GR, but it is not the case in our model since, unless $\alpha'=0$, $\nabla^{\mu} T_{\mu\nu}\neq0$ for a matter source with $w=-1/3$ (see  \eqref{eqn:w1db3} in Sec. \ref{sec:gensol} for the solution). Finally, in order to align with standard cosmology, we wish to avoid spatial curvature domination over dust in the early universe. This means that, using the continuity equation \eqref{cont1} for dust and the fact that $\rho_{k}\propto a^{-2}$, we must have \begin{align}  \label{domination1}
3 \left[\frac{\rho_{\rm m}+2 \alpha}{\rho_{\rm m}-2\alpha}\right]>2
\end{align}
leading to the following permitted interval 
\begin{align}
\label{curvt}
-\frac{\rho_{\rm m}}{10}<\alpha<\frac{\rho_{\rm m}}{2},
\end{align}
which is a tighter bound than the one given in \eqref{cond:dust}.

\subsection{Solving the continuity equation explicitly for $\rho(z)$}
\label{sec:gensol}

As mentioned in Sec. \ref{sec:EMLG}, one of the difficulties in studying EMSG type models is that it is usually not possible to obtain the explicit exact solution of $\rho$ in terms of scale factor $a$ (or redshift $z$). For instance, in \cite{Akarsu:2017ohj} which investigated cosmic acceleration in a dust only universe via EMPG, the explicit solution of $\rho_{\rm m}(z)$ could only be obtained through an approximation procedure. In this section, we investigate the cases providing explicit solutions of $\rho(z)$ and show that EMLG model \eqref{eqn:lnassumption} provides an exact solution for the dust only universe.

\begin{figure}[!bt]
\captionsetup{justification=raggedright,singlelinecheck=false,font=footnotesize}
\includegraphics[width=0.41\textwidth]{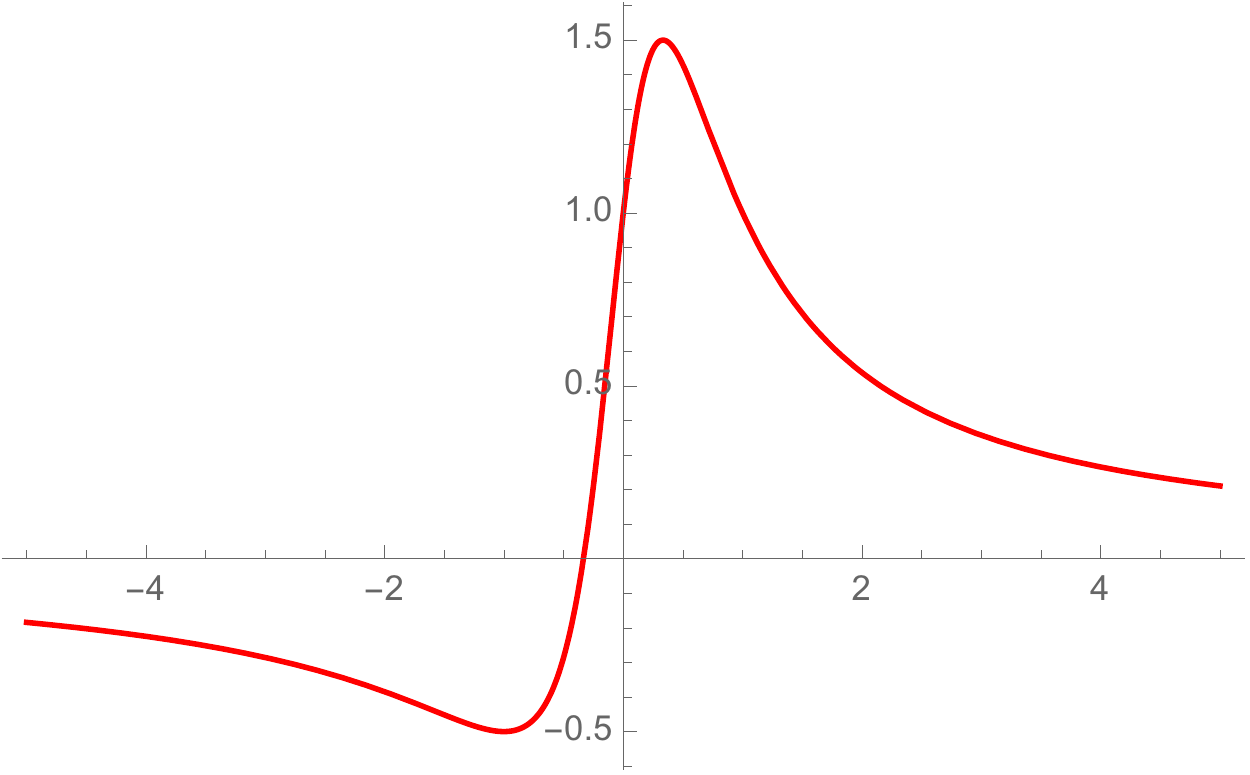}
\caption{The behaviour of the parameter $\beta$ (y-axis) for different equation of state parameters $w$ (x-axis), i.e., $\beta(w)$. The region of most interest has $-1\leq w\leq 1$.}
\label{beta_w}
\end{figure}

Defining 
\begin{equation}
\beta(w)=\frac{3w+1}{3w^2+1},
\end{equation} 
we rewrite \eqref{noncons} as
\begin{align}
\frac{\dot{\rho}}{\rho}\left[\frac{\rho-2\alpha}{\rho+2\alpha\beta}\right]=-3(1+w)\frac{\dot{a}}{a},
\end{align}
which can be solved implicitly as 
\begin{align} \label{imp_soln}
\rho\left(1+\frac{2\alpha}{\rho}\beta\right)^{\frac{1}{\beta}+1}\propto a^{-3(1+w)}.
\end{align}
We can then proceed by examining the behaviour of $\beta(w)$ plotted in Fig. \ref{beta_w}. 

We notice first that $\beta$ attains a maximum value of $3/2$ at $w=1/3$, and a minimum of $-1/2$ at $w=-1$; however, $\beta$ is not injective, and so there exist two values of $w$ that provide the same right-hand side of \eqref{imp_soln}. However, as the left hand side also has a $w$ dependence, the behaviour of our perfect fluid for the two equations of state will not coincide. 

At $w=-1/3$, we must note that $\beta=0$. At this point we consider the limiting behaviour of \eqref{imp_soln}, which takes exponential form:
\begin{align}
\label{eqn:w1db3}
\rho e^{\frac{2\alpha}{\rho}}\propto a^{-2}.
\end{align}
We could also recover this by integrating \eqref{noncons} directly with $w=-1/3$. This equation of state no longer corresponds to the behaviour of curvature terms as in GR, but describes the evolution of cosmic strings. We also note the similarities between the behaviour for $w=-1/3$ in this model and that in EMPG, as discussed in \cite{Board:2017ign}. However, we cannot solve the radiation dominated Universe explicitly.

This implicit solution \eqref{imp_soln} depends on the behaviour of the parameter $\beta$, and in general we would  not expect to find explicit solutions for the energy density in terms of the scale factor. In fact, we will be able to find explicit closed form solutions in certain physically relevant cases when \eqref{imp_soln} reduces to a polynomial in $\rho$ of degree at most four. If we write the exponent as $\frac{A}{B}=\frac{1}{\beta}+1$ as a fraction in its lowest terms ($A$, $B$ $\in \mathbb{Z}$, $B\neq 0$) we can determine the conditions on $A$ and $B$ such that the resulting equation is an appropriate polynomial. Once this is done, we can further constrain the exponent by considering the values which $\beta$ may take. It emerges that the only appropriate values that the exponent can take are integers in the list $\{-3, -2, -1, 2, 3, 4\}$. Two of these cases are of specific interest. The $-1$ case corresponds to $w=-1$, the equation of state for the conventional vacuum energy, in which case the exponent on the right hand side vanishes and we find that the energy density in this case, $\rho_{-1}$ is a constant, equal to its value today $\rho_{-1,0}$, that is: 
\begin{equation}
\rho_{-1}\equiv\rho_{-1,0}
\end{equation}
as in the GR case.

The second case of interest is $\beta=1$, in which case \eqref{imp_soln} reduces to a quadratic. This arises for the physically relevant cases of dust, $w=0$ and stiff fluid, $w=1$. This allows us to find an exact solution for the energy density in these cases, the specific form of which is discussed in the subsequent section. 

The remaining cases each result from a pair of values of $w$, but these values are irrational and thus unlikely to be of physical importance. Typically, one of the two values lies within the $-1<w<1$ range, and the other outside.

It is also important to note that although we have explicit solutions for these cases, and can examine features of \eqref{imp_soln} for others, we are not able to compare the behaviour of a single cosmological model using these solutions since they are each valid only for a single fluid Universe. In this study, we will investigate the late-time acceleration of the universe, accordingly, neglect the radiation and assume that there is only dust as the material source, for which, fortunately, EMLG provides us with explicit solution for $\rho(a)$. In Section \ref{radincld}, we will also briefly discuss possible analytical solutions of a Universe including radiation.

\subsection{Dust-filled Universe}
Since we will concentrate our discussions on the late-time acceleration of the universe, we assume that the radiation density is negligible, and the universe is spatially flat and filled only with dust. Accordingly, substituting $w=0$ and $k=0$ into the modified Friedmann equations \eqref{eq:rhoprime} and \eqref{eq:presprime}, they reduce to the following
\begin{equation}
3H^{2}= \rho_{\rm m}+\Lambda+\alpha'\rho_{\rm m,0}-\alpha'\rho_{\rm m,0} \ln\left(\rho_{\rm m}/\rho_{\rm m,0}\right) , \label{eq:rhoprime-w0}
\end{equation}
\begin{equation}
-2\dot{H}-3H^{2}=-\Lambda+ \alpha'\rho_{\rm m,0}\ln \left(\rho_{\rm m}/\rho_{\rm m,0}\right). \label{eq:presprime-w0}
\end{equation}
And for $w=0$, the continuity equation \eqref{noncons} is satisfied as
\begin{equation}
\dot{\rho}_{\rm m}+3H\rho_{\rm m} \left(\frac{\rho_{\rm m}+\alpha' \rho_{\rm m,0}}{\rho_{\rm m}-\alpha' \rho_{\rm m,0}}\right)=0,  \label{cont_dust}
\end{equation}
and hence as discussed above, we obtain the explicit solution
\begin{equation}
\begin{aligned}
\label{soln}
\rho_{\rm m}=&\frac{1}{2}\rho_{\rm m,0}(1+\alpha')^2 (1+z)^{3}-\alpha'\rho_{\rm m,0}  \\ 
&+\frac{1}{2}\rho_{\rm m,0}\sqrt{-4\alpha'^2+\left[(1+\alpha')^2(1+z)^{3}-2\alpha'\right]^2}\, ,   
\end{aligned}
\end{equation}
provided that $-1<\alpha'\leq1$, and using that $a=(1+z)^{-1}$. We note that as $\alpha'\rightarrow0$, in our solution $\rho_{\rm m}\rightarrow\rho_{\rm m,0} (1+z)^3$, the usual pressureless matter evolution, so we recover the standard $\Lambda$CDM model along with GR. We also note that \eqref{eq:rhoprime-w0} with $\Lambda=0$ at the present time reads $3H_{0}^{2}= \rho_{\rm m,0}+\alpha'\rho_{\rm m,0}$ and consequently $\Omega_{\rm m,0} (1+\alpha')=1$. Here we define the present day density parameters of dust and $\Lambda$ as $\Omega_{\rm m,0}=\frac{ \rho_{\rm m,0}}{3 H_{0}^2}$ and $\Omega_{\Lambda,0}=\frac{\Lambda}{3H_0^2}$. From the most recent observational results $\Omega_{\rm m,0}\approx0.3$ and therefore we estimate that $\alpha'\approx2.3$. However, our solution \eqref{soln} is not valid for this $\alpha'$ value. Thus, to be able to use the solution \eqref{soln}, we must include $\Lambda$ in our model, so that  \eqref{eq:rhoprime-w0}  implies that $\Omega_{\rm m,0} (1+\alpha')+\Omega_{\Lambda,0}=1$. We note that the intervals we deduced in Section \ref{sec:preliminary} for a viable cosmology are a subset of the interval needed for the validity of solution \eqref{soln} today. Namely, curvature domination discussion in \eqref{curvt} with the definition \eqref{eq:alphaprime} leads to a narrower interval for $\alpha'$. Considering that interval of $\alpha'$, $-0.20<\alpha'<1$, we find $1-2\;\Omega_{\rm m,0}<\Omega_{\Lambda,0}<1-0.8\;\Omega_{\rm m,0}$. Consequently, we estimate that the solution given in \eqref{soln} is valid for  $0.40\lesssim\Omega_{\Lambda,0}\lesssim0.76$. Furthermore, as $z\rightarrow-1$, the energy density $\rho\rightarrow-\alpha'\rho_{\rm m,0}=\rho_{\rm min}$. This means that if the universe were to expand forever, the energy density would never reach to zero. Instead there would be a minimum energy density limit as $\rho_{\rm min}=-\alpha'\rho_{\rm m,0}$, which in turn implies that $\alpha'$ must be negative in an  eternally expanding universe. Finally we note that the solution for equation of state $w=1$ is the same as the solution for dust, with $a\rightarrow a^2$.

\section{Improved $Om$ diagnostic of EMLG}
\label{sec:diag}
Cosmological models with late time acceleration, via DE in GR or modified gravity, can be examined with the use of null-diagnostics. One diagnostic is the jerk parameter $j=\frac{\dddot{a}}{aH^3}$, first introduced by Harrison \cite{har} (who denoted it by $Q$), which is simply equal to unity in $\Lambda$CDM (omitting radiation), $j_{\Lambda {\rm CDM}}=1$,  \cite{Sahni:2002fzA,Sahni:2002fzB,Visser:2003vq,Dunajski:2008tg}. Hence, any observational evidence which predicts a deviation from unity implies that late time acceleration is not due to the cosmological constant in GR. The second diagnostic is $Om(z)$ which is defined via an improved version in a recent study \cite{Sahni:2014dee} as follows:
\begin{equation} \label{def:Omh2}
Om h^2(z_i ; z_j)=\frac{h^2(z_i)-h^2(z_j)}{(1+z_i)^3-(1+z_j)^3},
\end{equation} 
where $h(z)=H(z)/100\,{\rm km\,s}^{-1}{\rm Mpc}^{-1}$ is the dimensionless reduced Hubble parameter. We note that $Om$ depends only on $H(z)$, and is therefore easier to determine from observations than $j$. Consequently, knowing the Hubble parameter at two or more redshifts, one can obtain the value of $Omh^2$ and conclude whether or not a dark energy modification to GR is the cosmological constant. In $\Lambda$CDM, omitting radiation (which is negligible in the late universe) we have
\begin{align}
h^2=h_0^2 \left[\Omega_{\rm m,0} (1+z)^3+1-\Omega_{\rm m,0}\right],
\end{align}
which simply gives a constant as
\begin{equation}
Om h^2(z_i ; z_j)=h_0^2 \Omega_{\rm m,0}.
\end{equation}

The estimates given in \cite{Sahni:2014dee} for the $Omh^2$ diagnostic consider $H(z_{1}=0) = 70.6 \pm 3.3 {\rm km\,s}^{-1}{\rm Mpc}^{-1}$ \cite{Efstathiou:2013via} based on the NGC 4258 maser distance, $H(z_{2}=0.57) = 92.4 \pm 4.5{\rm km\,s}^{-1}{\rm Mpc}^{-1}$ \cite{bao_2013} based on the clustering of galaxies in the SDSS-III BOSS DR9, and $H(z_{3}=2.34) = 222 \pm 7{\rm km\,s}^{-1}{\rm Mpc}^{-1}$ \cite{Delubac:2014aqe} based on the BAO in the Lyman-$\alpha$ forest of SDSS DR11 data and read
\begin{equation}
\begin{aligned}
Omh^2(z_{1};z_{2})=0.124 \pm 0.045, \\
Omh^2(z_{1};z_{3})=0.122 \pm 0.010, \\
Omh^2(z_{2};z_{3})=0.122 \pm 0.012. 
\end{aligned}
\end{equation}
Note that these model-independent values of $Omh^2$ for any two redshifts are stable at about $0.12$ which is in tension with, the value $Omh^2 = \Omega_{\rm m,0}h_0^2 = 0.1430 \pm 0.0011$ determined for the base $\Lambda$CDM model from the Planck 2018 release \cite{Aghanim:2018eyx}. Note that $Omh^2$ is not affected significantly by $H(z=0)$ (the accurate value of which is subject to a great debate in the contemporary cosmology) owing to the high-precision measurement of $H(z=2.34)$ \cite{Sahni:2014dee}.

It is argued in \cite{Sahni:2014dee} that this tension can be alleviated in models in which $\Lambda$ was dynamically screened in the past. In line with this, until Section \ref{obsresults}, we investigate the features of the EMLG model (parametrised by $\alpha'$) in comparison with the $\Lambda$CDM model mostly by referring to \cite{Sahni:2014dee}. Therefore, we intentionally make use of these three $H(z)$ data (rather than the latest data, which would not change our arguments in what follows) as well as the $\Omega_{\rm m,0}$ and $H_0$ values considered in \cite{Sahni:2014dee}. This allows us to demonstrate the effect of the EMLG model on $Omh^2$ diagnostics, with a properly chosen value for $\alpha'$, by a straightforward comparison with \cite{Sahni:2014dee}. We shall investigate the observational analyses of the EMLG model and compare with the $\Lambda$CDM model using the latest cosmological data in Section \ref{obsresults}.

\begin{figure}[t]
\captionsetup{justification=raggedright,singlelinecheck=false,font=footnotesize}
\includegraphics[width=0.41\textwidth]{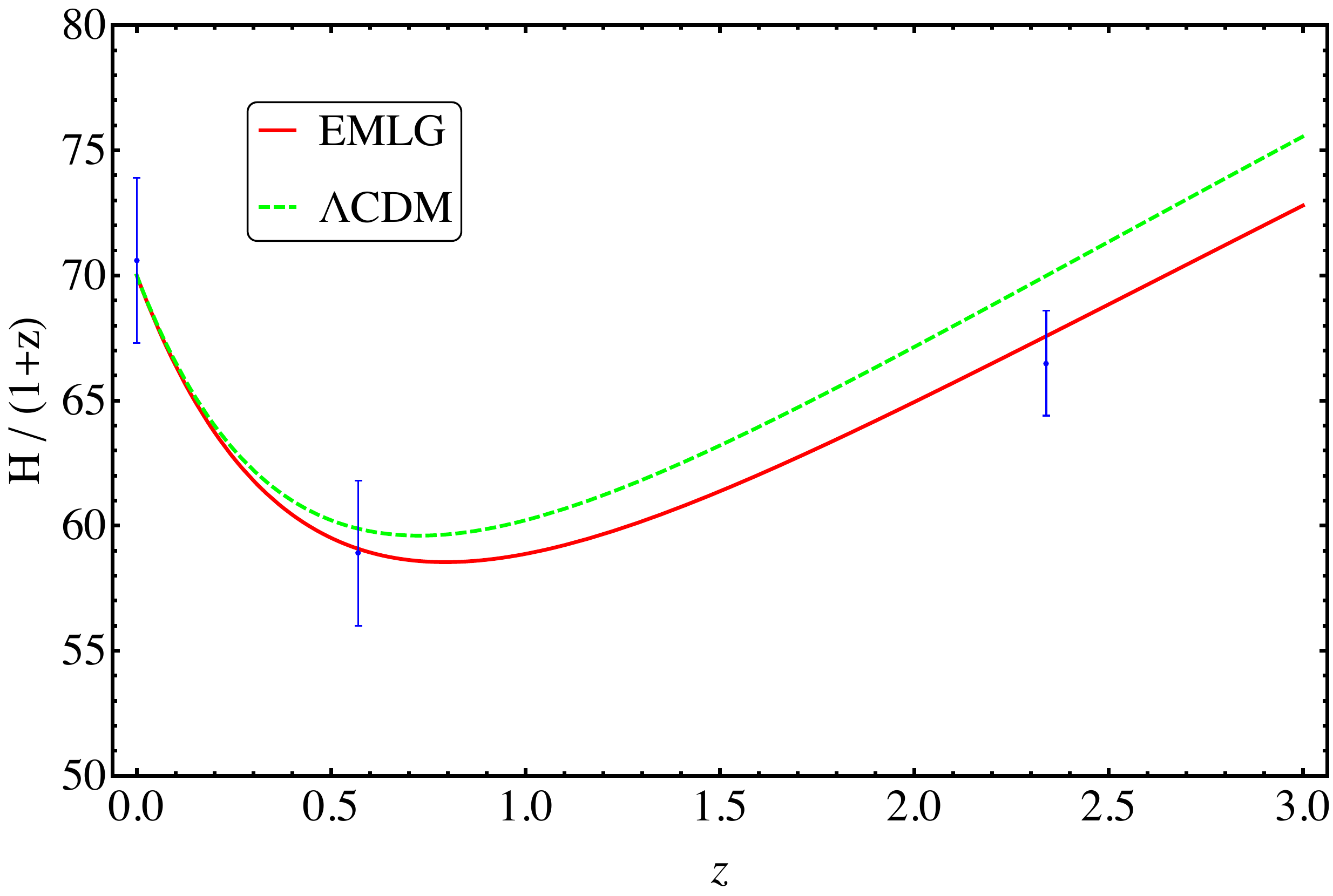}
\caption{$H(z)/(1+z)$ vs. $z$ graph of the EMLG and $\Lambda$CDM. Plotted by using $\Omega_{{\rm m},0}=0.28$, $H_{0}=70\,{\rm km\,s}^{-1}{\rm Mpc}^{-1}$ and $\alpha'=-0.04$. For the three observational $H(z)$ values with errors we consider those in \cite{Sahni:2014dee}.}
\label{H_div1+z_est}
\end{figure}

\subsection{EMLG cosmology in the light of null-diagnostics}
We now consider the $Om$ diagnostic expression defined in  \eqref{def:Omh2} for our model. Substituting the solution \eqref{soln} into \eqref{eq:rhoprime-w0}, we obtain
\begin{equation}
\begin{aligned}
h^2=&h_0^2\left\{1-\Omega_{\rm m,0}\left\{1-\frac{1}{2}\left[(1+\alpha')^2(1+z)^{3}-2\alpha'   \right.\right.\right.  \\ &\left.+\sqrt{-4\alpha'^2+\left[(1+\alpha')^2(1+z)^{3}-2\alpha'\right]^2}\right]   \\ &\left.+\alpha' \ln\left\{\frac{1}{2}\left[(1+\alpha')^2(1+z)^{3}-2\alpha'\right.\right.\right.  \\ &\left.\left.\left.\left.+\sqrt{-4\alpha'^2+\left[(1+\alpha')^2(1+z)^{3}-2\alpha'\right]^2}\right]\right\}\right\}\right\} \label{Hpar},
\end{aligned}
\end{equation}
where we use also the fact that $\Omega_{\Lambda,0}=1-(1+\alpha') \Omega_{\rm m,0}$. This leads to
\begin{equation}
\begin{aligned}\label{eqn:omh2EMLG}
Om h^2(z_i ;& z_j)=h_0^2 \Omega_{\rm m,0} \left\{\left(\alpha '+1\right)^2 \left(z_i+1\right){}^3\right.
 \\ &+\left.\sqrt{\left(\left(\alpha '+1\right)^2 \left(z_i+1\right){}^3-2 \alpha '\right){}^2-4 \alpha '^2}\right. \\ &\left.-2 \alpha ' \ln \left[\frac{1}{2} \left(-2 \alpha '+\left(\alpha '+1\right)^2 \left(z_i+1\right){}^3 \right.\right.\right. \\ &\left.\left.\left.+\sqrt{\left(\left(\alpha '+1\right)^2 \left(z_i+1\right){}^3-2 \alpha '\right){}^2-4 \alpha '^2}\right)\right]\right. \\ &\left.-\left(\alpha '+1\right)^2 \left(z_j+1\right){}^3\right.  \\ &\left.-\sqrt{\left(\left(\alpha '+1\right)^2 \left(z_j+1\right){}^3-2 \alpha '\right){}^2-4 \alpha '^2}\right. \\ &\left.+2 \alpha ' \ln \left[\frac{1}{2} \left(-2 \alpha '+\left(\alpha '+1\right)^2 \left(z_j+1\right){}^3\right.\right.\right. \\ &\left.\left.\left.+\sqrt{\left(\left(\alpha '+1\right)^2 \left(z_j+1\right){}^3-2 \alpha '\right){}^2-4 \alpha '^2}\right)\right]\right\}/  \\ &2 \left[\left(z_i+1\right){}^3-\left(z_j+1\right){}^3\right].
\end{aligned}
\end{equation}
Following the three $H(z)$ data given in \cite{Sahni:2014dee}, in Fig. \ref{H_div1+z_est}, we plot $H(z)/(1+z)$ with respect to redshift using $\Omega_{\rm m,0}=0.28$ and $H_{0}=70{\rm km\,s}^{-1}{\rm Mpc}^{-1}$ for both the $\Lambda$CDM model (green) and the EMLG model with $\alpha'=-0.04$ (red), which provides us $H(z)/(1+z)$ in agreement with all data points whereas the one for $\Lambda$CDM does not fit to the data point from $z=2.34$. The true constant of the model in the action \eqref{eq:action} is, accordingly, $\alpha=-0.02\rho_{\rm m,0}$. The model-independent value of the $Om$ diagnostic estimated in \cite{Sahni:2014dee} is quite stable at $Om h^2\simeq0.12$ and is in tension with the $\Lambda$CDM-based value $Om h^2({\Lambda}$CDM) $\simeq0.14$. On the other hand, for the EMLG model with $\Omega_{\rm m,0}=0.28$, $H_{0}=70{\rm km\,s}^{-1}{\rm Mpc}^{-1}$ and $\alpha'=-0.04$, we find $Om h^2(z_1 ; z_2)=0.129$, $Om h^2(z_1 ; z_3)=0.127$ and $Om h^2(z_2 ; z_3)=0.127$ where $z_1=0$, $z_2=0.57$ and $z_3=2.34$. Note that these are in good agreement with the estimates given in \cite{Sahni:2014dee}.


\subsection{A comparison via general relativistic interpretation}

In \cite{Sahni:2014dee}, it is suggested that lower values for $Om h^2$ can be obtained in models in which the cosmological constant was screened by a dynamically evolving counter-term $f(z)$ in the past. Accordingly, $H^2(z)$ is modified, with respect to the  $\Lambda$CDM model, as
\begin{align}
H^2(z)=\frac{1}{3} \rho_{\rm m,0} (1+z)^3+\frac{\Lambda}{3}-f(z).  \label{Sahni9}
\end{align}
and at a redshift $z_*$, \;$\Lambda/3$ is balanced by $f(z)$ (i.e. $f(z_*)=\Lambda/3$). Comparing \eqref{Sahni9} and \eqref{eq:rhoprime-w0}, along with our solution given in \eqref{soln}, it emerges that in our model
\begin{equation}
\begin{aligned}
f(z)&=\frac{1}{6} \rho_{\rm m,0}\left[\left(2-(1+\alpha')^2\right) (1+z)^3\right]   \\  &-\frac{1}{6} \rho_{\rm m,0}\sqrt{-4\alpha'^2+\left[(1+\alpha')^2(1+z)^{3}-2\alpha'\right]^2}    \\ &+\frac{1}{3} \rho_{\rm m,0} \alpha' \ln\left\{\frac{1}{2}\left[(1+\alpha')^2(1+z)^{3}-2\alpha'\right.\right.    \\ &\left.\left.+\sqrt{-4\alpha'^2+\left[(1+\alpha')^2(1+z)^{3}-2\alpha'\right]^2}\right]\right\}. \label{f_model}
\end{aligned}
\end{equation}
It is not possible to calculate the redshift, $z_{*}$, exactly from \eqref{f_model}. However, for $\Omega_{\rm m,0}=0.28$ and $\alpha'=-0.04$, we can numerically calculate that $z_{*}=2.29$ for our model (similar to the value $z_{*}\simeq 2.4$ given in \cite{Sahni:2014dee}). 

Furthermore, \cite{Sahni:2014dee} suggests that evolving DE models in which $\Lambda$, as part of the dark energy, was screened in the past provide a better fit for the BAO data than the $\Lambda$CDM model, as well as alleviating the tension discussed in the preceding two sections. It is also noted that in such evolving DE models, the effective EoS of the DE displays a pole at high redshifts. A pole in $w_{\rm DE}$ implies that the energy density of the DE changes sign at that redshift value. This behavior of the DE is also discussed in another study \cite{Aubourg:2014yra} by the BOSS collaboration using the BBAO, SN and Planck data sets. In the next section, we will investigate the EMLG model from this perspective.

\subsection{Effective dynamical dark energy}  \label{dark energy}

In order to test our model in light of the above discussion, we reconstruct the model by defining an effective DE by rewriting \eqref{eq:rhoprime-w0} and \eqref{eq:presprime-w0} in the following form:
\begin{align}  \label{rho_darkenergy}
3H^{2}&= \rho_{\rm m,0} (1+z)^3+ \rho_{\rm DE}, \\ -2\dot{H}-3H^{2}&= p_{\rm DE}.
\end{align}
Thus, the energy density and pressure of the effective DE are given by
\begin{equation}
\begin{aligned}
\rho_{\rm DE}=&\rho_{\rm m}+\alpha' \rho_{\rm m,0}\left[1-\ln\left(\rho_{\rm m}/\rho_{\rm m,0}\right)\right]\\
&- \rho_{\rm m,0} (1+z)^3+\Lambda,
\label{rhoDE}
\end{aligned}
\end{equation}
\begin{equation}
p_{\rm DE}=\alpha' \rho_{\rm m,0}\ln \left(\rho_{\rm m}/\rho_{\rm m,0}\right)-\Lambda.
\end{equation}
Next, using \eqref{soln} in these equations we obtain $\rho_{\rm DE}$ and $p_{\rm DE}$ as follows;
\begin{align}
\rho_{\rm DE}=&\frac{1}{2}\rho_{\rm m,0}\left\{\left[(1+\alpha')^2-2\right] (1+z)^3\right.  \nonumber \\ 
&+ \left.\sqrt{-4\alpha'^2+\left[(1+\alpha')^2(1+z)^{3}-2\alpha'\right]^2}\right\} \nonumber \\  
&-\alpha'\rho_{\rm m,0}\ln\left\{\frac{1}{2}\left[(1+\alpha')^2(1+z)^{3}-2\alpha'\right.\right. \nonumber \\ 
&+\left.\left.\sqrt{-4\alpha'^2+\left[(1+\alpha')^2(1+z)^{3}-2\alpha'\right]^2}\right]\right\}+\Lambda, \label{eq:rhoDE}
\end{align}
\begin{align}
p_{\rm DE}=&\alpha' \rho_{\rm m,0}\ln\left\{\frac{1}{2}\left[(1+\alpha')^2(1+z)^{3}-2\alpha'\right.\right. \nonumber \\ 
&+\left.\left.\sqrt{-4\alpha'^2+\left[(1+\alpha')^2(1+z)^{3}-2\alpha'\right]^2}\right]\right\}-\Lambda.
\end{align}  
The corresponding EoS parameter $w_{\rm DE}=\frac{p_{\rm DE}}{\rho_{\rm DE}}$ is
\begin{align}  
w_{\rm DE}=&-1+\left\{\rho_{\rm m}-\rho_{\rm m,0}(1+z)^3+ \alpha' \rho_{\rm m,0}\right\}/ \nonumber \\  \big\{&\rho_{\rm m}-\rho_{\rm m,0}(1+z)^3 \nonumber \\ &+\alpha' \rho_{\rm m,0}\left[1-\ln\left(\rho_{\rm m}/\rho_{\rm m,0}\right)\right]+\Lambda\big\}.
 \label{eosDE}
\end{align}
Defining the density parameter of the effective dark energy for today as $\Omega_{\rm DE,0}=\frac{\rho_{\rm DE,0}}{3 H_{0}^2}$,  \eqref{eosDE} together with \eqref{soln} and \eqref{rhoDE} gives
\begin{equation}
\begin{aligned}
w_{\rm  DE}=&-1+(1-\Omega_{\rm DE,0})\bigg[\left((1+\alpha')^2-2\right) (1+z)^3  \\ &+\sqrt{-4\alpha'^2+\left[(1+\alpha')^2(1+z)^{3}-2\alpha'\right]^2}\bigg]\bigg/ \\&\bigg\{(1-\Omega_{\rm DE,0})\bigg\{\left[(1+\alpha')^2-2\right] (1+z)^3-2\alpha' \\ &+\sqrt{-4\alpha'^2+\left[(1+\alpha')^2(1+z)^{3}-2\alpha'\right]^2} \\ &-2\alpha'\ln\bigg(\frac{1}{2}\bigg[(1+\alpha')^2(1+z)^{3}-2\alpha' \\ &+\sqrt{-4\alpha'^2+\left[(1+\alpha')^2(1+z)^{3}-2\alpha'\right]^2}\bigg]\bigg)\bigg\} \\
&+2\Omega_{\rm DE,0}\bigg\},        \label{w_DE}
\end{aligned}
\end{equation}
where we have used the fact that $\Omega_{\rm m,0}+\Omega_{\rm DE,0}=1$. The present-day value of the EoS parameter of the effective DE is
\begin{equation}
w_{\rm DE,0}=-1+\alpha'\,\frac{1-\Omega_{\rm DE,0}}{\Omega_{\rm DE,0}}.
\end{equation}
We note that it lies in the `phantom' region  ($w<-1$) for $\alpha'<0$. Specifically, $w_{\rm DE,0}=-1.0156$ for $\alpha'=-0.04$ and $\Omega_{\rm DE,0}=0.72$. 


\begin{figure}[t!!]
\captionsetup{justification=raggedright,singlelinecheck=false,font=footnotesize}
\includegraphics[width=0.41\textwidth]{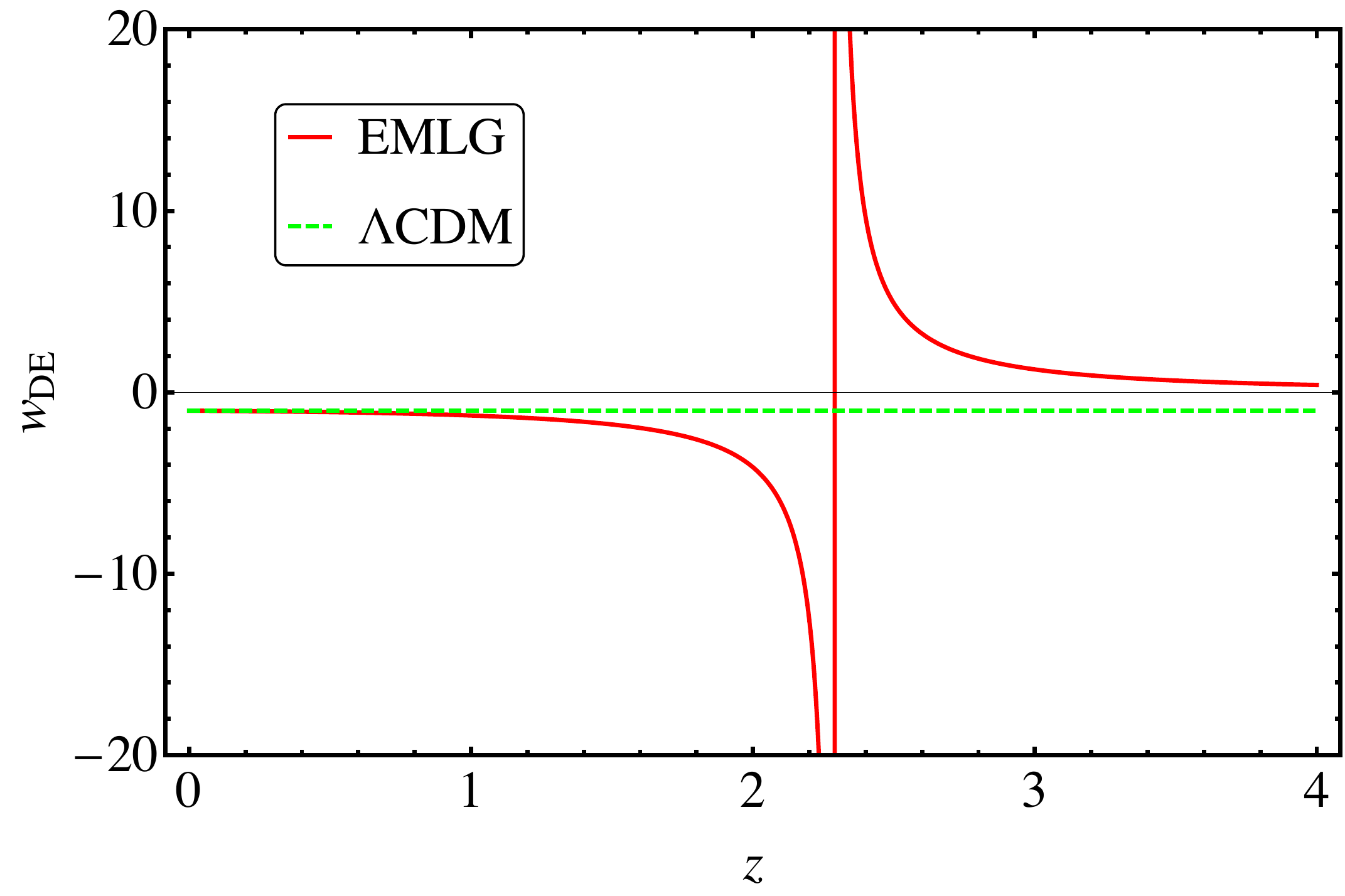}
\caption{$w_{\rm DE}$ versus $z$ graphs of the EMLG and $\Lambda$CDM. Plotted by using $\Omega_{\rm m,0}=0.28$ and $\alpha'=-0.04$. $|w_{\rm DE}|\rightarrow\infty$ at $z=2.29$ in EMLG. }
 \label{wDEvsz}
\end{figure}

\begin{figure}[t!!]
\captionsetup{justification=raggedright,singlelinecheck=false,font=footnotesize}
\includegraphics[width=0.41\textwidth]{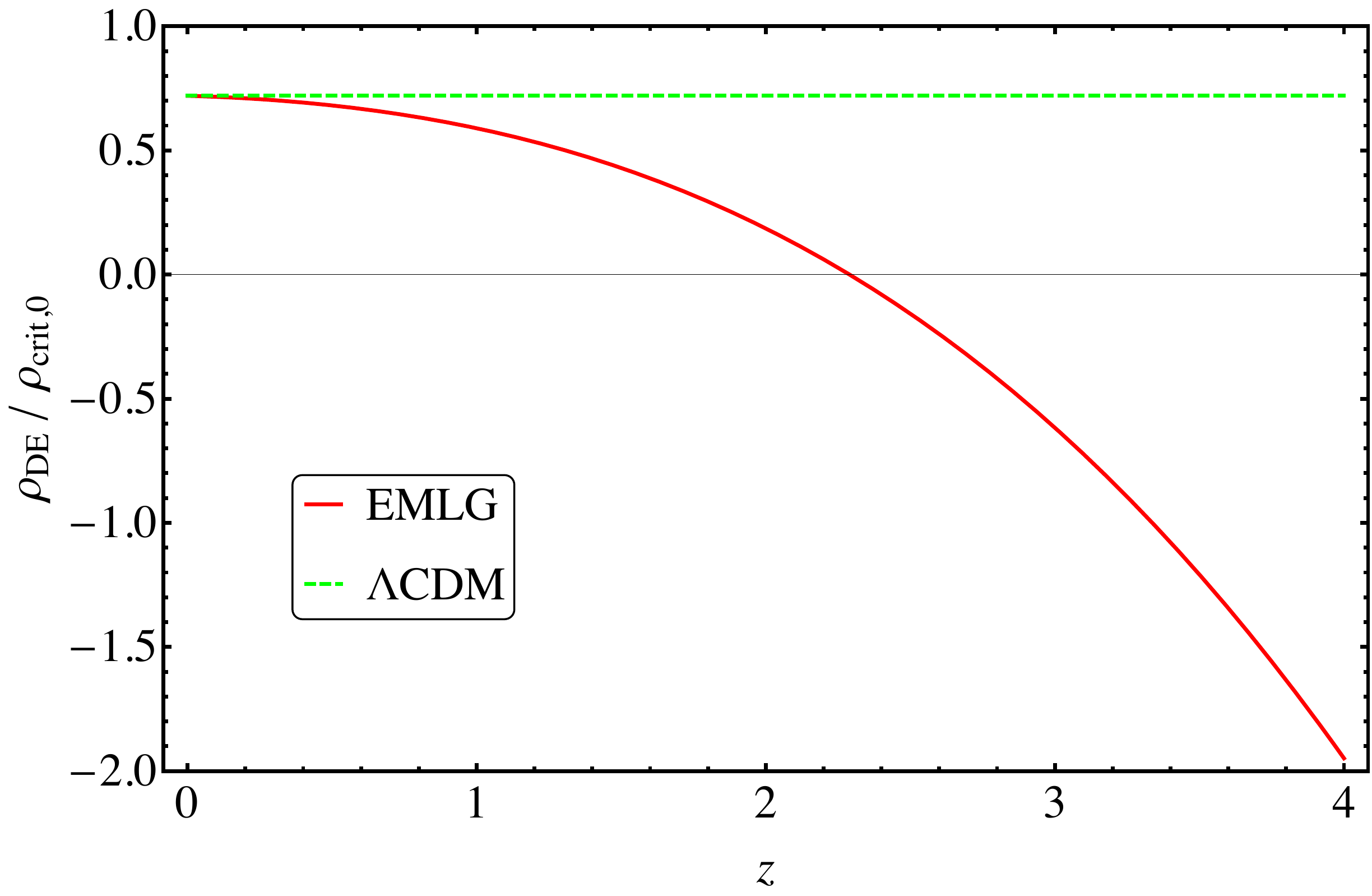}
\caption{$\rho_{\rm DE}/\rho_{\rm crit,0}$ versus $z$ graphs of the EMLG and $\Lambda$CDM. Plotted by using $\Omega_{{\rm m},0}=0.28$ and $\alpha'=-0.04$. $\rho_{\rm DE}=0$ at $z=2.29$ in EMLG.}
 \label{rhoDEvsz}
\end{figure}

\begin{figure}[!bt]
\captionsetup{justification=raggedright,singlelinecheck=false,font=footnotesize}
\includegraphics[width=0.41\textwidth]{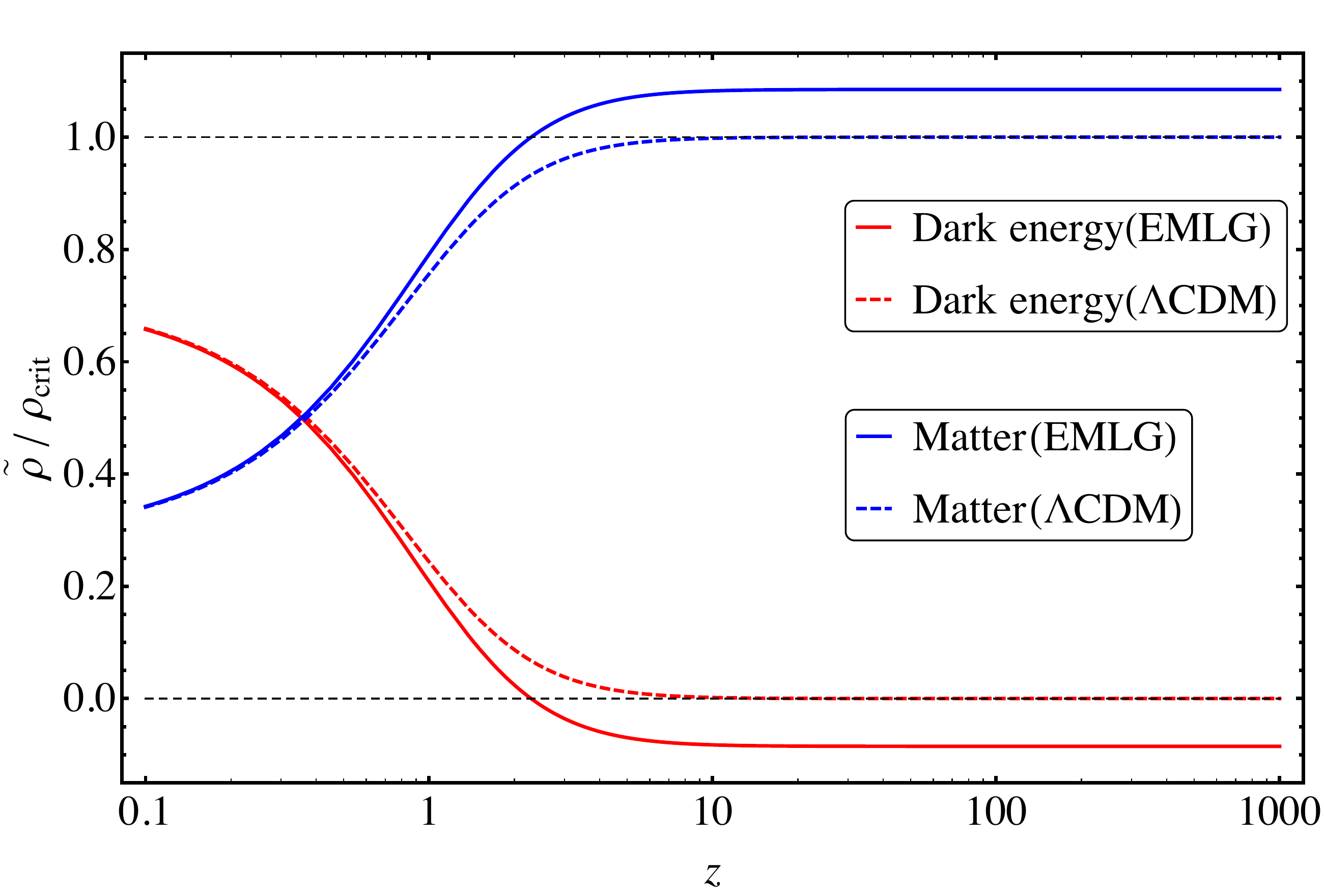}
\caption{Density parameters (shown as $\tilde{\rho}/\rho_{\rm crit}$) vs. $z$ graphs of the EMLG and $\Lambda$CDM for dust and effective dark energy. Here $\tilde{\rho}=\rho_{\rm m,0} (1+z)^3$ for matter and $\tilde{\rho}=\rho_{\rm DE}$ for effective dark energy. Plotted by using $\Omega_{{\rm m},0}=0.28$ and $\alpha'=-0.04$.} 
\label{rhoDEdivH2}
\end{figure}

As may be seen from  \eqref{w_DE}, the model reduces to $\Lambda$CDM for $\alpha'=0$ giving $w_{\rm DE}=w_{\rm DE,0}=-1$. We now plot illustrative figures by using $\Omega_{{\rm m},0}=0.28$ and $\alpha'=-0.04$. With these values, we see from  \eqref{eq:rhoDE} that $\rho_{\rm DE}=0$ at $z=2.29$. In accordance with the arguments in \cite{Sahni:2014dee}, within the effective DE source interpretation of our model, $\Lambda$ is screened at the redshift $z_{*}=2.29$ and the effective EoS of the DE exhibits a pole at the same redshift (which is very similar to the estimate $z_{*}\simeq 2.4$ made in \cite{Sahni:2014dee}). We depict the pole of $w_{\rm DE}$ at $z=2.29$ in Fig.\ref{wDEvsz}, which is due to $\rho_{\rm DE}$ changing sign at that redshift, as can be seen from Fig.\ref{rhoDEvsz}. Note that Fig.\ref{rhoDEvsz} shows clearly that the sign change at $z=2.29$ is in agreement with Fig.11 of \cite{Aubourg:2014yra} revealing that $\rho_{\rm DE}$ passes below zero at a redshift in the interval $1.6\leq z\leq 3.0$. We also display, both for the EMLG and $\Lambda$CDM models, the density parameters of dust, $\Omega_{\rm m}=\rho_{\rm m}/3H^2$, and the effective DE, $\Omega_{\rm DE}=\rho_{\rm DE}/3H^2$, ($\Omega_{\Lambda}=\rho_{\Lambda}/3H^2$ for the $\Lambda$CDM model) up to $z=1100$ in Fig. \ref{rhoDEdivH2}. Note that the density parameters are the same for $z=0$ and do not differ much for low redshifts. For large redshifts, in contrast, the unusual behavior of the EMLG model emerges, so that $\Omega_{\rm m}$ becomes equal to unity at $z=z_{*}=2.29$ (at $z\rightarrow\infty$ for the $\Lambda$CDM model) and then settles in a plateau larger than unity for $z>z_{*}=2.29$, which results from $\rho_{\rm DE}$ becoming negative at $z=z_*=2.29$.

\begin{figure}[!ht]
\captionsetup{justification=raggedright,singlelinecheck=false,font=footnotesize}
\includegraphics[width=0.41\textwidth]{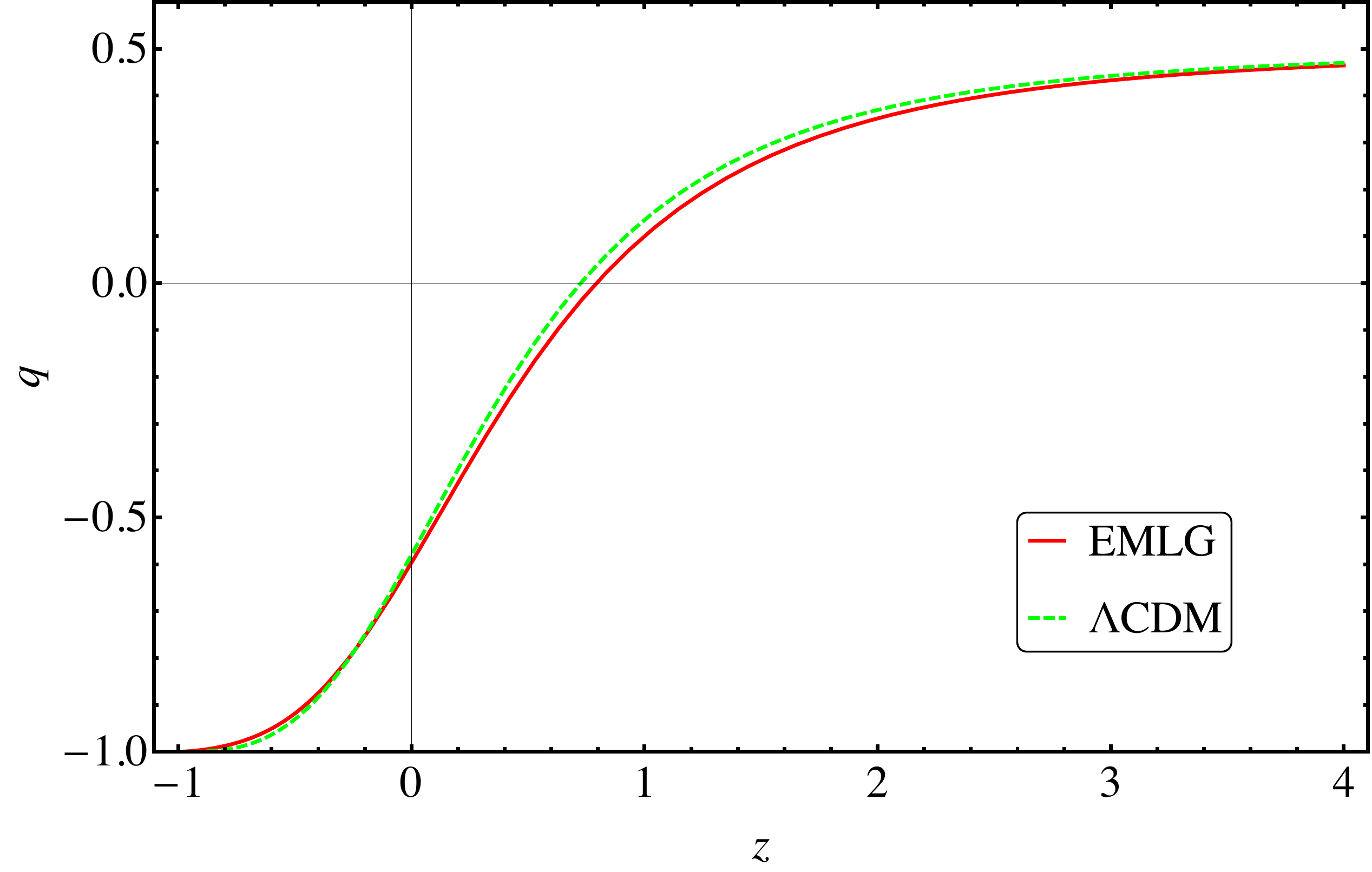}
\includegraphics[width=0.41\textwidth]{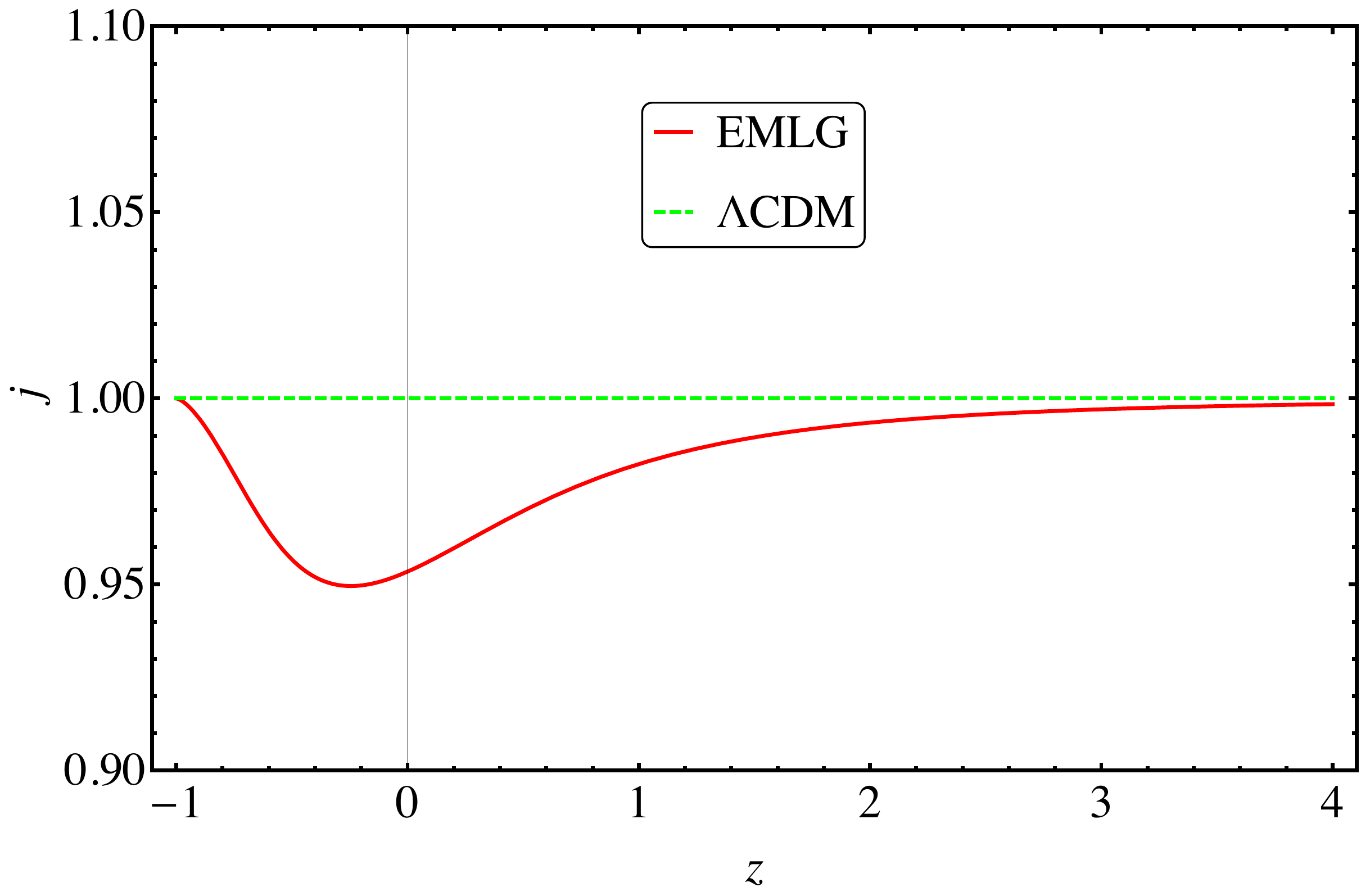}
\caption{$q(z)$ vs. $z$ \textbf{(upper panel)} and $j(z)$ vs. $z$ \textbf{(lower panel)} graphs of the EMLG and $\Lambda$CDM. Plotted by using $\Omega_{{\rm m},0}=0.28$, $H_{0}=70\,{\rm km\,s}^{-1}{\rm Mpc}^{-1}$ and $\alpha'=-0.04$.} 
\label{qvsz}
\end{figure}

Next we calculate two important kinematical parameters that are of interest in cosmology in order to compare different models. Firstly, we calculate the deceleration parameter, $q=-1-\frac{\dot{H}}{H^2}$, as
\begin{equation}
q=-1+\frac{3}{2}\; \frac{\Omega_{\rm m,0} \left[(\rho_{\rm m}/\rho_{\rm m,0})+\alpha'\right]}{1-\Omega_{\rm m,0}\left[1-(\rho_{\rm m}/\rho_{\rm m,0})+\alpha' \ln(\rho_{\rm m}/\rho_{m,0})\right]},  \label{decpar}
\end{equation}
which can be written in terms of redshift, by using \eqref{soln}, as
\begin{equation}
\begin{aligned}
q=&-1+\frac{3}{4} \Omega_{\rm m,0}\bigg[(1+\alpha')^2(1+z)^{3}  \\&+\sqrt{-4\alpha'^2+\left[(1+\alpha')^2(1+z)^{3}-2\alpha'\right]^2}\bigg]\bigg/   \\   &\bigg\{1-\Omega_{\rm m,0}\bigg\{1-\frac{1}{2} \big[(1+\alpha')^2(1+z)^{3}-2\alpha' \\ &+\sqrt{-4\alpha'^2+\left[(1+\alpha')^2(1+z)^{3}-2\alpha'\right]^2}\bigg]\bigg\} \\ &+\alpha'\ln\bigg\{\frac{1}{2}\big[(1+\alpha')^2(1+z)^{3}-2\alpha' \\ &+\sqrt{-4\alpha'^2+\left[(1+\alpha')^2(1+z)^{3}-2\alpha'\right]^2}\bigg]\bigg\}\bigg\}.
\end{aligned}
\end{equation}
Setting $\alpha'=0$ recovers the expression for these parameters in $\Lambda$CDM. We note that $q\rightarrow-1$ as $z\rightarrow-1$, implying that our model asymptotically approaches $\Lambda$CDM in the far future. For large redshifts, $z\gg1$, in \eqref{decpar} the deceleration parameter of the dust dominated era in $\Lambda$CDM, $q=1/2$, is recovered. Calculating the current value of the deceleration parameter, we find $q_{0}=-1+\frac{3}{2} \Omega_{\rm m,0}(1+\alpha')$. As can be seen in the top panel of Fig.\ref{qvsz}, the accelerated expansion begins at $z_{\rm tr}\approx0.79$ and the present time value of the deceleration parameter is $q_0=-0.60$, whereas these are $z_{\rm tr}\approx 0.73$ and $q_{0}=-0.58$ for $\Lambda$CDM model. Secondly, we calculate the jerk parameter $j=\frac{\dddot{a}}{aH^3}$, which was discussed in Sec. \ref{sec:diag} and, as mentioned, is simply equal to unity for $\Lambda$CDM (ommiting radiation). In contrast, for EMLG $j$ is dynamical and is given by
\begin{equation}
\begin{aligned}
j=&\bigg\{\alpha' \rho_{0} \Omega_{0}  (1+z)^2 \rho_z^2-\alpha' \rho_{0} \Omega_{0}  (1+z) \rho \big[(1+z) \rho_{zz} \\ &-2 \rho_{z}\big]+\rho^2 \big[\Omega_{0}  (1+z) \big((1+z) \rho_{zz}-2 \rho_{z}\big) \\ &-2 \rho_{0} \big(\alpha'\Omega_{0}  \ln \left(\rho/\rho_{0}\right)+\Omega_{0} -1\big)\big]+2 \Omega_{0}  \rho^3\bigg\}\bigg/  \\ &\bigg\{2 \rho^2 \big[\Omega_{0}  \rho-\rho_{0} \big(\alpha'  \Omega_{0}  \ln \left(\rho/\rho_{0}\right)+\Omega_{0} -1\big)\big]\bigg\},
\end{aligned}
\end{equation}
where we have written $\rho=\rho_{\rm m}(z)$, $\Omega_0=\Omega_{m,0}$, and a subscript of $z$ denotes differentiation with respect to redshift. The explicit expression in terms of redshift can be obtained by substituting $\rho_{\rm m}(z)$ from \eqref{soln}, which we do not provide explicitly for reasons of brevity. $j(z)$ is then depicted in the lower panel of Fig.\ref{qvsz} which illustrates the dynamical nature of the jerk parameter in EMLG. It deviates from unity at $z\sim0$ but we have $j\rightarrow 1$ in both limits as either $z\rightarrow\infty$ or $z\rightarrow-1$, hence EMLG recovers the kinematics of $\Lambda$CDM both at early times, and in the far future.


\subsection{Screening of $\Lambda$ by the non-conservation of dust}
\label{sec:secreen}

In Section \ref{dark energy}, we rearranged the original field equations of the EMLG model, \eqref{eq:rhoprime-w0} and \eqref{eq:presprime-w0}, in order to compare with the model first described in \cite{Sahni:2014dee}. For this comparison, we assumed that the energy density of the dust behaves as in GR, $\rho_{\rm m}\propto (1+z)^3$, and then compensated it as a part of the effective DE \eqref{rho_darkenergy}. In other words, we assume that all of the terms with $\alpha'$, including those coming from the true matter energy density \eqref{soln} of EMLG, contribute to the energy density of the effective DE. Through this  comparison, we have determined the parameter of our model, $\alpha'$, with which EMLG relaxes the issues of the $\Lambda$CDM model stated in \cite{Sahni:2014dee}.

\begin{figure}[!bt]   
\captionsetup{justification=raggedright,singlelinecheck=false,font=footnotesize}
\includegraphics[width=0.41\textwidth]{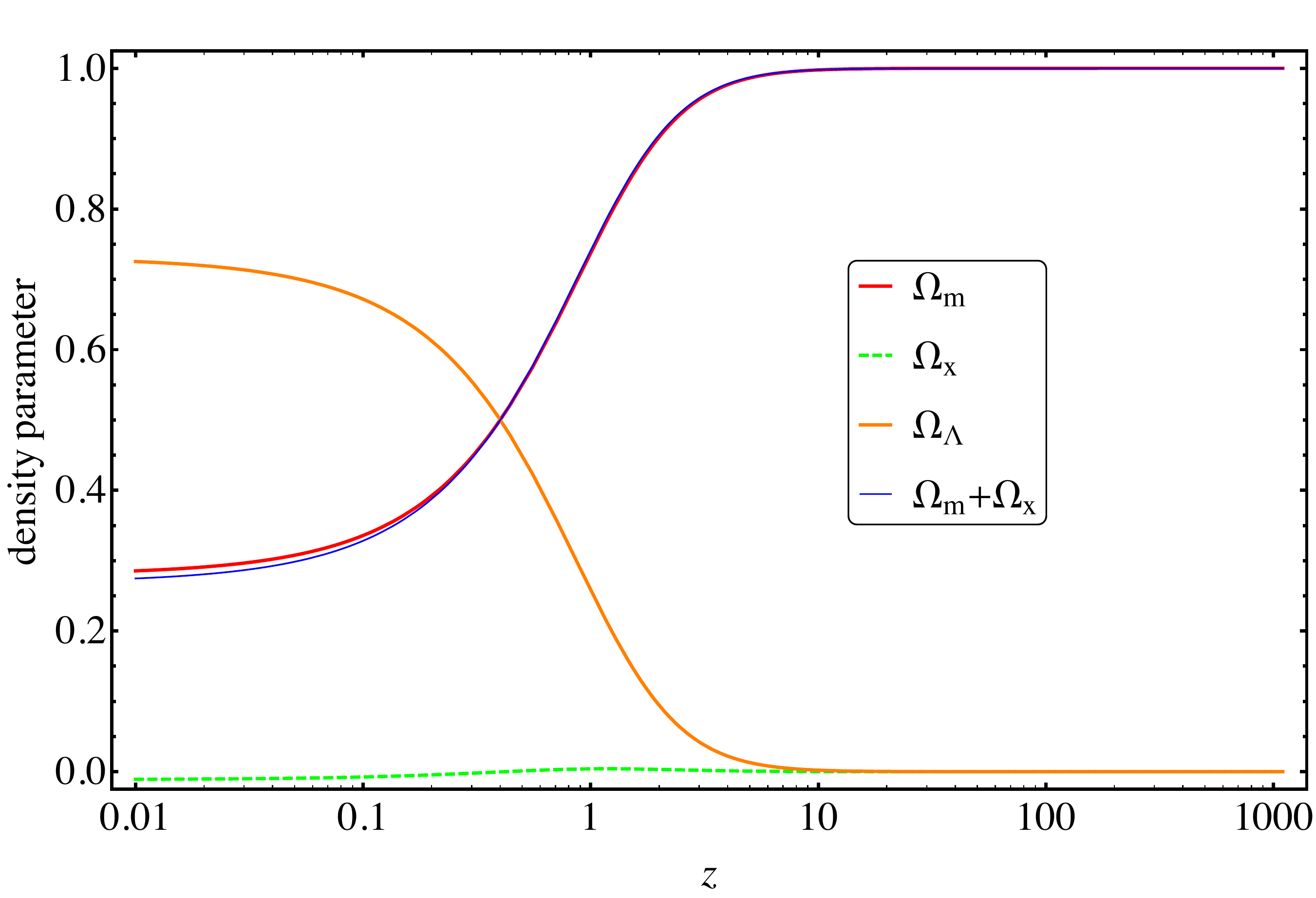}
\caption{$\Omega$ vs. $z$ graphs of the EMLG for matter ($\Omega_{\rm m}$), modification terms ($\Omega_{x}$), cosmological constant ($\Omega_{\Lambda}$) and matter+modification ($\Omega_{m}+\Omega_{x}$). Plotted by using $\Omega_{{\rm m},0}=0.28$, $H_{0}=70\,{\rm km\,s}^{-1}{\rm Mpc}^{-1}$ and $\alpha'=-0.04$.} 
\label{Omega_EMLG}
\end{figure}

We now examine the actual behavior of dust in EMLG. The energy density of dust in EMLG is given by \eqref{soln} and includes terms with the EMLG modification parameter $\alpha'$. Furthermore, we have new terms with $\alpha'$ in the original field equations, \eqref{eq:rhoprime-w0} and \eqref{eq:presprime-w0}, arising due to the EMLG modification to  GR. As a result, both the energy density of dust and the forms of the energy density and pressure equations of our model differ from those of GR. Consequently, we find it useful to depict, in Fig. \ref{Omega_EMLG}, the redshift dependency of the density parameters corresponding to the components of the energy density equation \eqref{eq:rhoprime-w0}. To do so, we define $\Omega_{\rm m}=\rho_{\rm m}/3H^2$ (red) for dust, $\Omega_{\Lambda}=\Lambda/3H^2$ (yellow) for $\Lambda$ and $\Omega_{\rm X}=[\alpha'\rho_{\rm m,0}-\alpha'\rho_{\rm m,0} \ln\left(\rho_{\rm m}/\rho_{\rm m,0}\right)]/3H^2$ (green) for the new terms which arise due to the EMLG modification. We use $\Omega_{{\rm m},0}=0.28$, $H_{0}=70\,{\rm km\,s}^{-1}{\rm Mpc}^{-1}$ and $\alpha'=-0.04$, the same values used in previous sections. We note the small and non-monotonic contribution from $\Omega_{\rm X}$ in \eqref{eq:rhoprime-w0}.

\begin{figure}[!bt]   
\captionsetup{justification=raggedright,singlelinecheck=false,font=footnotesize}
\includegraphics[width=0.41\textwidth]{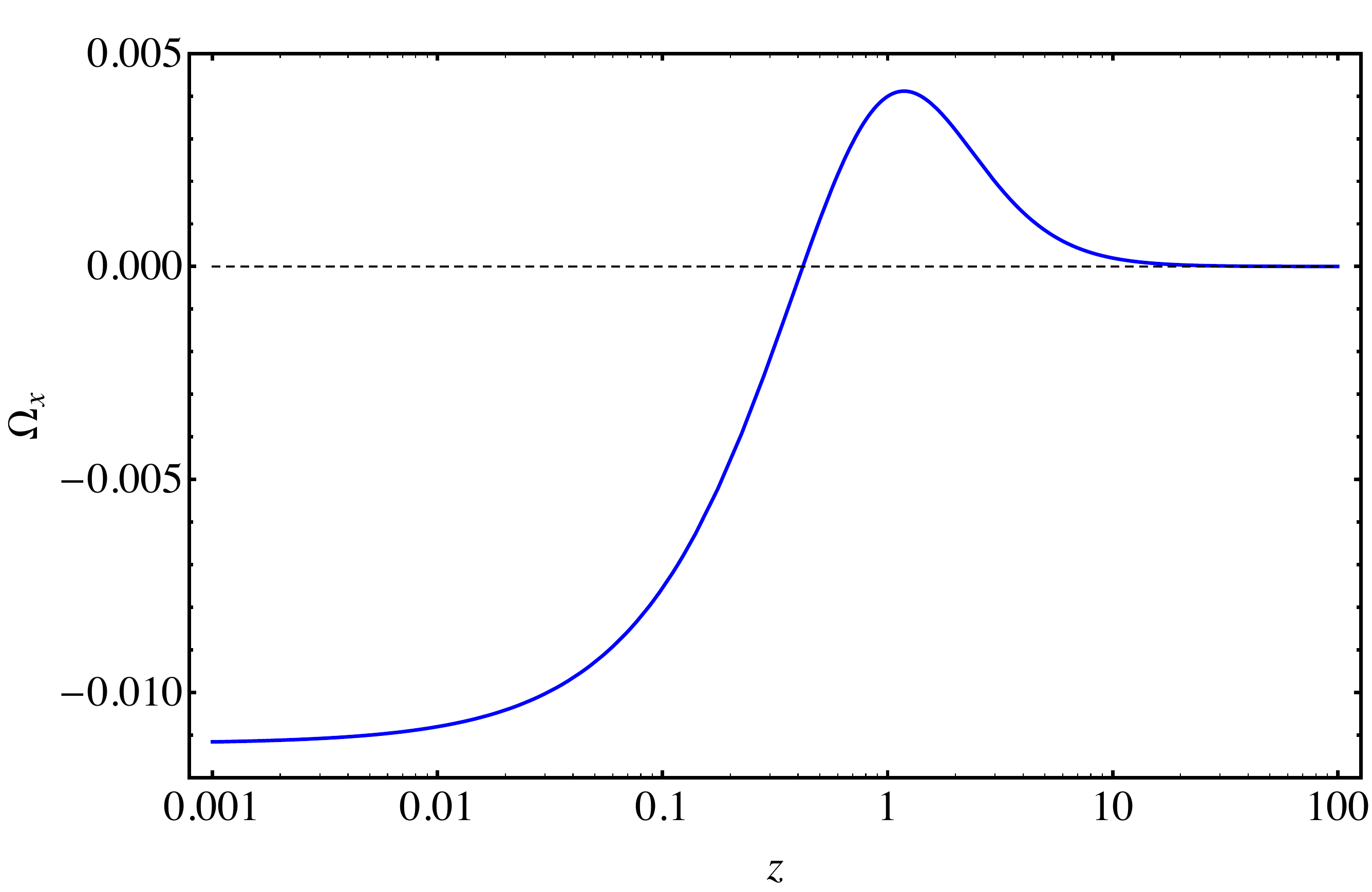}
\caption{The density parameter of modification terms ($\Omega_{x}$) vs. $z$ graph of the EMLG. Plotted by using $\Omega_{{\rm m},0}=0.28$, $H_{0}=70\,{\rm km\,s}^{-1}{\rm Mpc}^{-1}$ and $\alpha'=-0.04$.} 
\label{OmegaX_EMLG}
\end{figure}

For a better view, we depict $\Omega_{\rm X}(z)$ separately in Fig. \ref{OmegaX_EMLG}. This figure is of particular interest since it reveals an important point about the model under consideration; that the contribution from $\Omega_{\rm X}$ is negative at low redshifts, positive at $z\sim 1$ and then, whilst remaining positive, asymptotically approaches zero at larger redshifts. This means that $\Omega_{\rm X}$, due to the EMLG modification, screens $\Lambda$ only at low redshifts in contrast to the arguments given in \cite{Sahni:2014dee}. On the other hand, within the effective DE source interpretation of our model in line with \cite{Sahni:2014dee,Aubourg:2014yra}, we have already shown that $\rho_{\rm DE}$ is positive at low redshifts and passes below zero at $z=2.29$ exactly as suggested in \cite{Sahni:2014dee,Aubourg:2014yra}. This implies that the feature of screening $\Lambda$ in the EMLG model does not arise from the new type of contributions of dust on the right-hand side of \eqref{eq:rhoprime-w0} which appear as an effective source with constant inertial mass density as $\rho'+p'=\alpha'\rho_{\rm m,0}$ (see \ref{sec:cimd}), but instead from the altered redshift dependency of $\rho_{\rm m}$ due to the non-conservation of the EMT in the EMLG model. 

\subsection{Inclusion of radiation}
\label{radincld}
In order to investigate the implications of our model for the early universe while preserving its agreement with the current data for the late universe, we need to look for solutions in the case that radiation is the second source besides dust. Including both fluids as sources in our model results in complicated field equations including the cross terms of $\rho_{\rm r}$ and $\rho_{\rm m}$ which make exact solutions impossible. On the other hand, if we use the same $\alpha'=-0.04$ value which corresponds to $\alpha=-0.02\rho_{m,0}$ for radiation, it remains outside today's viability interval \eqref{int-rad} as we know from observations that $\rho_{m,0}/\rho_{r,0}\sim10^{3}$. This arises from the fact that the interval \eqref{int-rad} is valid only for a mono-fluid universe. We would need to decrease the absolute value $|\alpha|$ to find viable cosmological solutions when our model contains radiation as well. However, this would result in compromising the goodness of fit of our model with the latest data compared to that of $\Lambda$CDM for the late-time accelerated expansion of the universe. Thus, we conclude that it does not seem possible to expand our model by both adding radiation and preserving the features we have been discussing so far when there is only one $\alpha$ parameter involving in both sources.

A recent study \cite{Akarsu:2018aro} shows that different sources can couple to gravity in different ways for a particular example of $f(T_{\mu\nu}T^{\mu\nu})$ modification. One can follow the same idea in EMLG. Namely, the model can be constructed using different $\alpha$ parameters for different types of sources which means that different gravitational couplings occur for each source. To do so, one can start with a modification term as follows
\begin{align}
\label{sumf}
f(T_{\mu\nu}T^{\mu\nu})=\sum_{i}\alpha_{i}\ln(\lambda_{i}\, T_{\mu\nu}^{(i)}T^{\mu\nu}_{(i)}),
\end{align}
where $\alpha_{i}$ (the coupling parameter) and $\lambda_{i}$ are the constants for $i^{th}$ fluid. Note that the sum over $i$ in \eqref{sumf} evades the issue of cross terms occurring in the case of more than one fluid. However, the number of free parameters is increased. To relax this issue, fluids can be separated as conventional sources, such as radiation ($\gamma$, $\nu$) or baryons (b), and dark sector/unknown sources like cold dark matter. Then, one can assume that known sources couple to gravity according to GR, that is the corresponding $\alpha_{i}$'s are zero, whilst dark sector/unknown sources couple in accordance with the modified theory \cite{Akarsu:2018aro}. With this idea, the field equations in EMLG read
\begin{equation}
\begin{aligned}
3H^{2}=&\Lambda+\rho_{\gamma}+\rho_{b}+ \rho_{\rm cdm}\\
&+\alpha'\rho_{\rm cdm,0}\left[1- \ln\left(\frac{\rho_{\rm cdm}}{\rho_{\rm cdm,0}}\right)\right] ,
\end{aligned}
\end{equation}
\begin{equation}
-2\dot{H}-3H^{2}= -\Lambda+\frac{\rho_{\gamma}}{3}+\alpha'\rho_{\rm cdm,0}\ln\left(\frac{\rho_{\rm cdm}}{\rho_{\rm cdm,0}}\right).
\end{equation}
Here $\rho_{\gamma}\propto(1+z)^4$, $\rho_{b}\propto(1+z)^3$ as in GR and $\rho_{\rm cdm}$ obeys the modified continuity equation \eqref{cont1} when $w=0$, which gives the energy density solution in \eqref{soln}. We reserve such an investigation to our future works.

\section{Constraints from latest cosmological data}
\label{obsresults}

In the preceding sections we have investigated theoretically the EMLG model, particularly in comparison with the studies \cite{Sahni:2014dee,Aubourg:2014yra}. For convenience, we assumed  the values of the Hubble constant and dust density parameter as used in \cite{Sahni:2014dee} and took a value of the coupling parameter of the EMLG modification so as to produce results similar to those discussed in \cite{Sahni:2014dee}. In this section we analyse the constraints on the parameters of the EMLG model from the latest observational data and discuss the model further. In order to explore the parameter space, we make use of a modified version of a simple and fast Markov Chain Monte Carlo (MCMC) code, named SimpleMC \cite{Anze,Aubourg:2014yra}, that computes expansion rates and distances using the Friedmann equation. The code uses a compressed version of a recent reanalysis of Type Ia supernova (SN) data, and high-precision Baryon Acoustic Oscillation measurements (BAO) at different redshifts with $z<2.36$ \cite{Aubourg:2014yra}. We also include a collection of currently available $H(z)$ measurements (CC), see \cite{Gomez-Valent:2018hwc} and references therein. For an extended review of cosmological parameter inference see \cite{EPadilla}. Table \ref{tab:priors} displays the parameters used throughout this paper along with the corresponding flat priors. Note that we do not consider CMB data in our analysis, because the current EMLG model does not contain radiation (see Section \ref{radincld} for the relevant discussion) and therefore we avoid radiation in the $\Lambda$CDM model in order to be able to compare these two models under the same conditions. 

We use the dimensionless Hubble parameter $h=H/100\, {\rm km\,s}^{-1}{\rm Mpc}^{-1}$ \cite{WMAP11a}, the physical baryon density $\Omega_b h^2$ and the pressureless matter density (including CDM) $\Omega_{\rm m}$. Throughout the analysis we assume flat priors over our sampling parameters: $\Omega_{{\rm m},0}=[0.05,1.5]$ for the pressureless matter density parameter today, $\Omega_{{\rm b},0} h_0^2=[0.02,0.025]$ for the baryon density parameter today and $h_0=[0.4,1.0]$ for the reduced Hubble constant. For the EMLG parameter we assume $\alpha' =[-1,1]$, which is also the validity interval of our solution, see  \eqref{soln}.

\begin{figure}[h!]
\captionsetup{justification=raggedright,singlelinecheck=false,font=footnotesize}
\par
\begin{center}         
\includegraphics[trim = 1mm  1mm 1mm 1mm, clip, width=9.0cm, height=9.0cm]{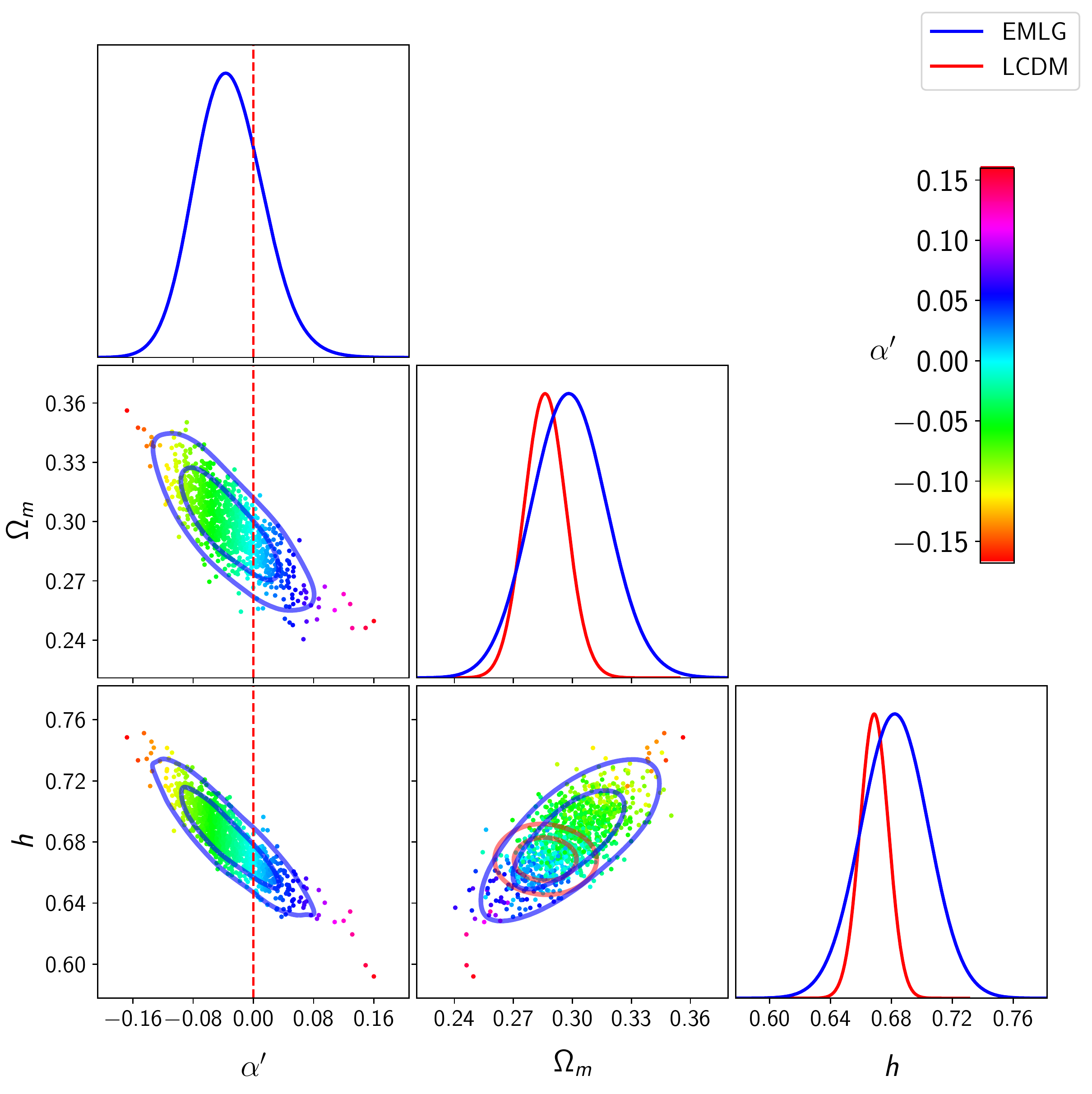}        
\end{center}
\caption{1D and 2D marginalized posterior distributions of the parameters used to describe the EMLG model (blue) and the $\Lambda$CDM model (red). Scatter points indicate values of $\alpha'$ labelled by the colour bar, and the vertical line corresponds to the $\Lambda$CDM case ($\alpha'=0$).}
\label{fig:constraints}
\end{figure}

For simplicity, and noticing the near-gaussianity of the posterior 
distributions (Fig. \ref{fig:constraints}), to perform a model selection we include the Akaike Information Criterion (AIC) \cite{Akaike74}, defined as:
\begin{eqnarray}
    {\rm AIC }= -2 \ln \mathcal{L}_{\rm max} + 2K,
\end{eqnarray}
where the first term incorporates the goodness-of-fit through the likelihood $\mathcal{L}$, and the second term is interpreted as the penalisation factor given by two times the number of parameters ($K$) of the model. The preferred model is then the one that minimises ${\rm AIC}$. A rule of thumb used in the literature is that if the AIC value of a model relative to that of the preferred model $\Delta\rm AIC\leq2$, it has substantial support; if $4\leq\Delta\rm AIC\leq7$, it has considerably less support, with respect to the preferred model. A Bayesian model selection applied to the dark-energy equation of state is performed by \cite{JAVazquez, Hee:2016ce, Tamayo:2019gqj}.
\begin{table}[t!]
\footnotesize
\captionsetup{justification=raggedright,singlelinecheck=false,font=footnotesize}
\caption{Constraints on the EMLG parameters using the combined datasets BAO+SN+CC. For one-tailed distributions the upper limit 95\% CL is given. For two-tailed the 68\% is shown. Parameters and ranges of the uniform priors assumed in our analysis. 
Derived parameters are labeled with~$^*$.}
\begin{tabular}{cccc}
\cline{1-4}\noalign{\smallskip}
\vspace{0.15cm}
Parameter   &    $\quad\quad$ EMLG $\quad\quad$ & $\quad\quad$ $\Lambda$CDM $\quad\quad$  & $\quad\quad$ Priors $\quad\quad$   \\
 \hline
\vspace{0.15cm}
$\Omega_{\rm m,0}$		& $0.2983 \pm 0.0185$ & $0.2861\pm 0.0102 $	& [0.05,1.5] 	\\
\vspace{0.15cm}
$\Omega_{b,0}h_0^2$			& $0.02196 \pm 0.00045$ & $0.02205\pm 0.00045$	 &[0.02, 0.025]\\
\vspace{0.15cm}
$h_0$					& $0.682 \pm 0.021$ &	 $0.668 \pm 0.009$   	&   [0.4, 1.0] 	\\
\vspace{0.15cm}
$\alpha'$				& $-0.032\pm 0.043$			&	 [0] 	& [-1, 1]\\
\hline
\vspace{0.15cm}
$^* w_{\rm DE,0}$		& $-1.015\pm 0.019$		&	 [-1] 	&	\\
\vspace{0.15cm}
$^*z_*$					& $2.23\pm0.81$		&	 - 	&	\\

\hline
\vspace{0.15cm}
$-\ln \mathcal{L_{\rm max}}$				&  34.22 	&	 34.49	& --  \\
AIC		&  76.44 	&	 74.98	& --  \\
\hline
\label{tab:priors}
\end{tabular}
\end{table}

Table \ref{tab:priors} summarizes the observational constraints on the free parameters
(as well as the derived parameters, labelled by $*$) of the EMLG model using the combined dataset BAO+SN+CC. For comparison, we also include parameters describing the $\Lambda$CDM model. 
We notice the EMLG model fits the data slightly better, however EMLG is penalized by the inclusion of the extra parameter $\alpha$, viz., with $\Delta\rm AIC=1.46$, and hence it has evidence to be a good model w.r.t. the $\Lambda$CDM model, but the $\Lambda$CDM model is slightly preferred over it. Figure \ref{fig:constraints} displays the 1D and 2D marginalized posterior distributions of the parameters used to describe the EMLG model (blue) and the $\Lambda$CDM model (red). The inner ellipses show the 68\% confidence region, and the outer edges the 95\% region. Scatter points indicate values of $\alpha'$ labelled by the colour bar, and the  vertical line corresponds to the $\Lambda$CDM case ($\alpha'=0$).

The data constrains the parameter of the EMLG model as $\alpha'= -0.032\pm 0.043$ at 68 $\%$ C.L., which well covers $\alpha'=0$ ($\Lambda$CDM), but prefers slightly negative values. In comparison with the $\Lambda$CDM model ($\alpha'=0$), the preference of the EMLG model for slightly negative values of $\alpha'$ leads to a widening of the 1D posterior distributions of $\Omega_{\rm m,0}$ and $h_0$ towards larger values, which in turn shifts the peak values of both parameters to larger values as well. Indeed, we see in Table \ref{tab:priors} that, in comparison with $\Lambda$CDM, the EMLG model predicts larger $\Omega_{\rm m,0}$ and $h_0$ values along with larger errors against the data. The strong anti-correlations on the parameters $\Omega_{\rm m,0}$ and $\alpha'$ and also on the $h_0$ and $\alpha'$ observed in 2D marginalised posterior distributions for the EMLG are an interesting point to note. These two anti-correlations lead to a correlation on the parameters $\Omega_{\rm m,0}$ and $h_0$, so that the larger negative values of $\alpha'$ lead to larger values of both of them. In contrast, in $\Lambda$CDM there is no noticeable correlation on the parameters $\Omega_{\rm m,0}$ and $h_0$. These can be observed directly in the $\{\Omega_{\rm m,0},h_0\}$ panel of the 3D scatter color Fig.\ref{fig:constraints}. For the EMLG model, 2D $\{\Omega_{\rm m,0},h_0\}$ contours exhibit a tilt of about 45 degrees and the more reddish (implying larger negative values of $\alpha'$) corresponds to larger $\Omega_{\rm m,0}$ and $h_0$ values.  

We study the constraints on the $Om h^2(z_i ; z_j)$ diagnostic values of the EMLG model using \eqref{eqn:omh2EMLG} for $\{z_1,z_2,z_3\}=\{0,0.57,2.34\}$, where the latter two redshift values are chosen in accordance with the BOSS CMASS and Lyman-$\alpha$ forest measurements of $H(z)$, and obtain
\begin{equation}
\begin{aligned}
Om h^2(z_1 ; z_2)&=0.132 \pm 0.008, \\
Om h^2(z_1 ; z_3)&=0.130 \pm 0.006, \quad\textnormal{(EMLG)}\\
Om h^2(z_2 ; z_3)&=0.130 \pm 0.006. 
\end{aligned}
\end{equation}
Using the $\Omega_{\rm m,0}$ and $h_0$ obtained for the EMLG model in $Om h^2(z_i ; z_j)=\Omega_{\rm m,0}h_0^2$ of the $\Lambda$CDM model (assuming $\alpha'=0$) we find a larger value as $Om h^2(z_i ; z_j)=0.139\pm0.012$, which clearly shows the reducing effect of $\alpha'<0$ on the $Om h^2(z_i ; z_j)$. On the other hand, for the $\Lambda$CDM model, in our analysis the data predicts a slightly lower value, with respect to those in the EMLG model, as
\begin{equation}
Om  h^2(z_i ; z_j)=0.128 \pm 0.006,\quad (\Lambda{\rm CDM})
\end{equation}
which results from $h_0=0.668\pm0.009$ and $\Omega_{\rm m,0}=0.2861\pm0.0102$. Note that this low value for the $\Lambda$CDM model is very much consistent with $Omh^2 \approx 0.122 \pm 0.010$ from BOSS CMASS and Lyman-$\alpha$ forest measurements of $H(z)$, which is obtained since we do not consider CMB data in our analysis. Indeed, the Planck 2018 \cite{Aghanim:2018eyx} release gives $\Omega_{\rm m,0}h_0^2 = 0.1430 \pm 0.0011$ from $h_0=0.674\pm0.005$ and $\Omega_{\rm m,0}=0.315\pm0.007$.
This shows that reducing the value of $Om h^2$ in $\Lambda$CDM comes at the cost of reducing $\Omega_{\rm m,0}$ to values in tension with the Planck result, and also of reducing $h_0$ to values which, whilst consistent with Planck results, exacerbate the persistent tension in the measurement of $H_0$ between the Planck $\Lambda$CDM model and direct measurements from astrophysical data.

\begin{figure}[!bt]
\captionsetup{justification=raggedright,singlelinecheck=false,font=footnotesize}
\includegraphics[trim = 1mm  1mm 1mm 1mm, clip, width=2.6cm, height=3.cm]{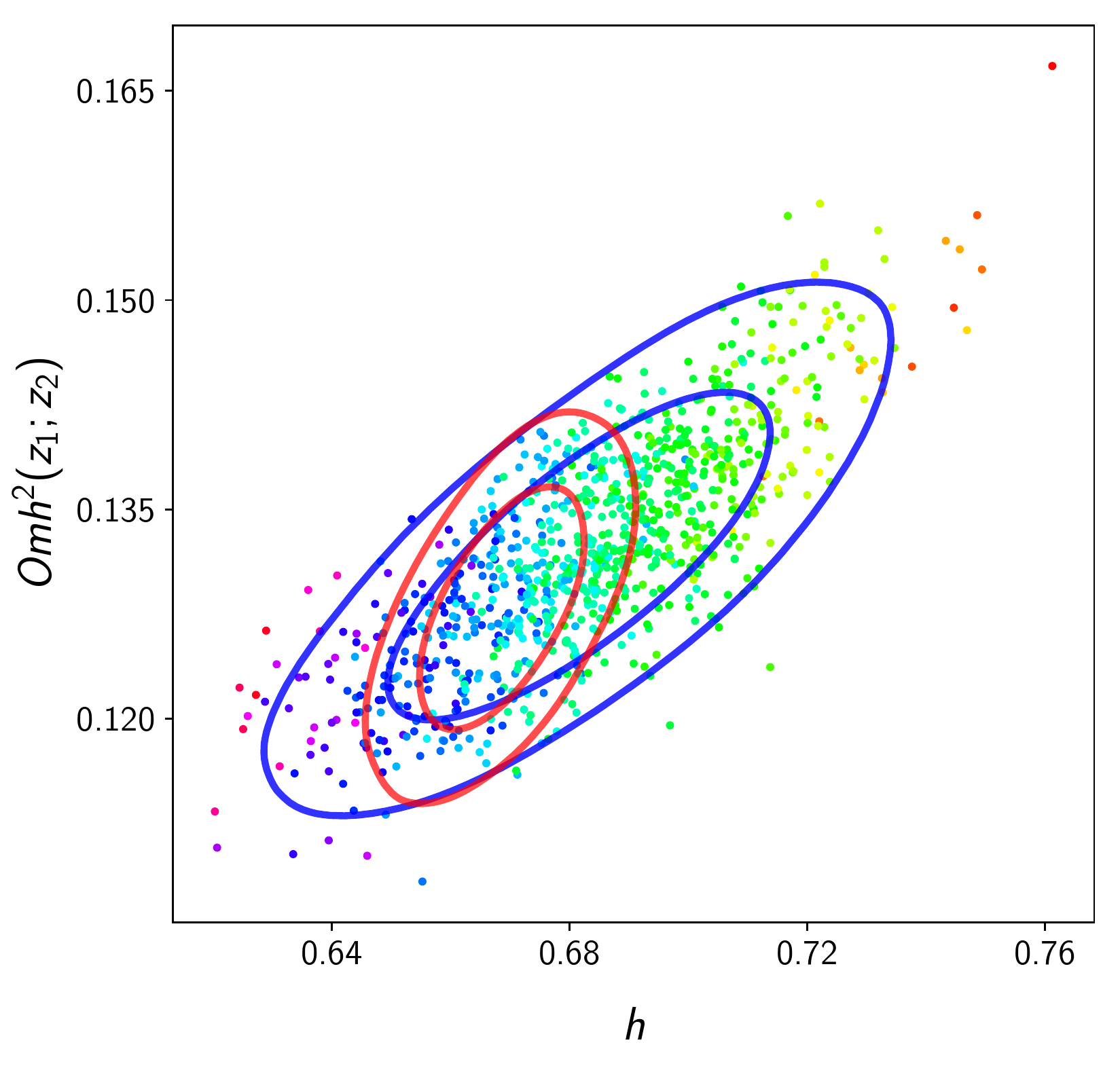}
\includegraphics[trim = 1mm  1mm 1mm 1mm, clip, width=2.6cm, height=3.cm]{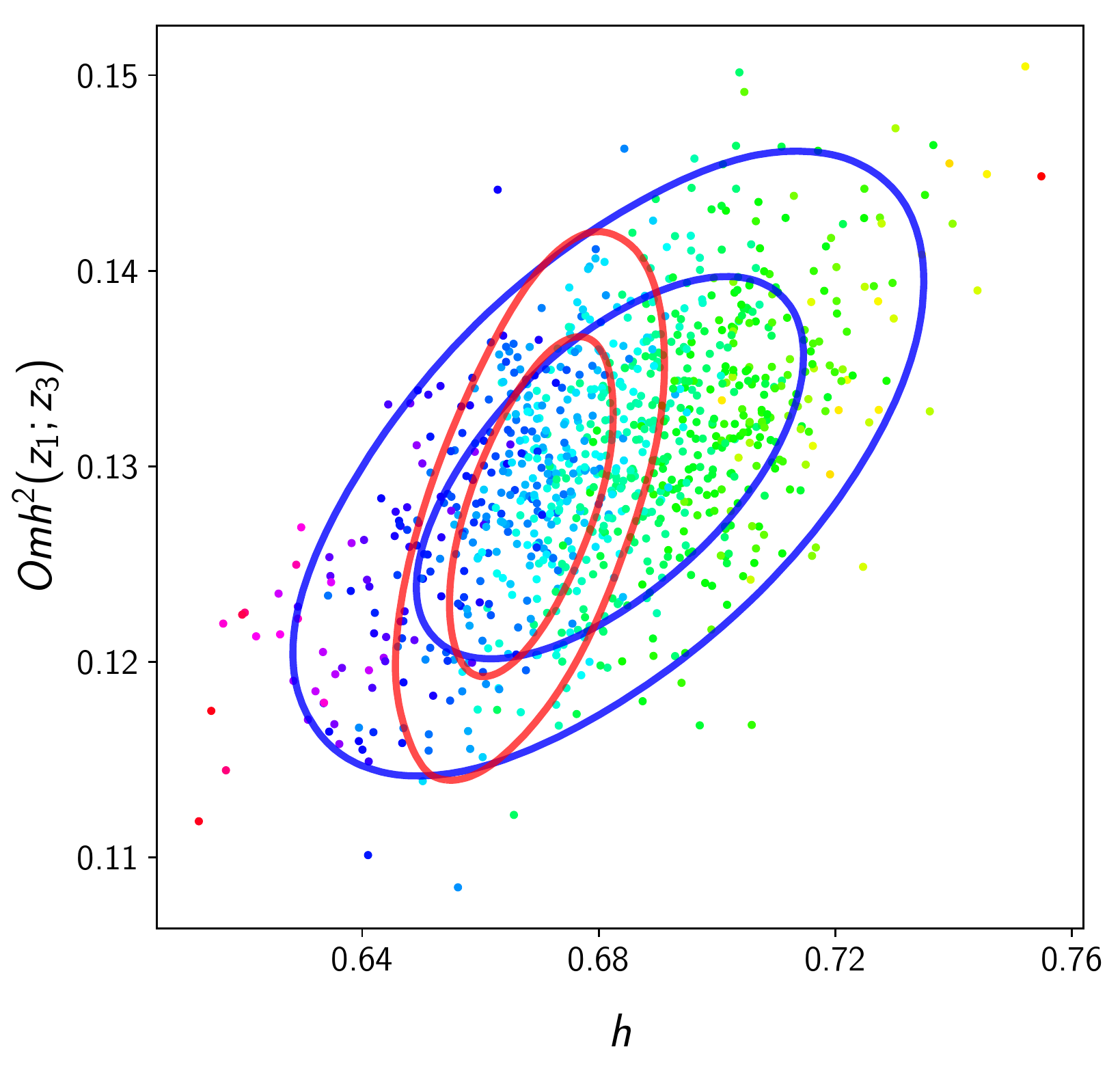}
\includegraphics[trim = 1mm  1mm 1mm 1mm, clip, width=2.6cm, height=3.cm]{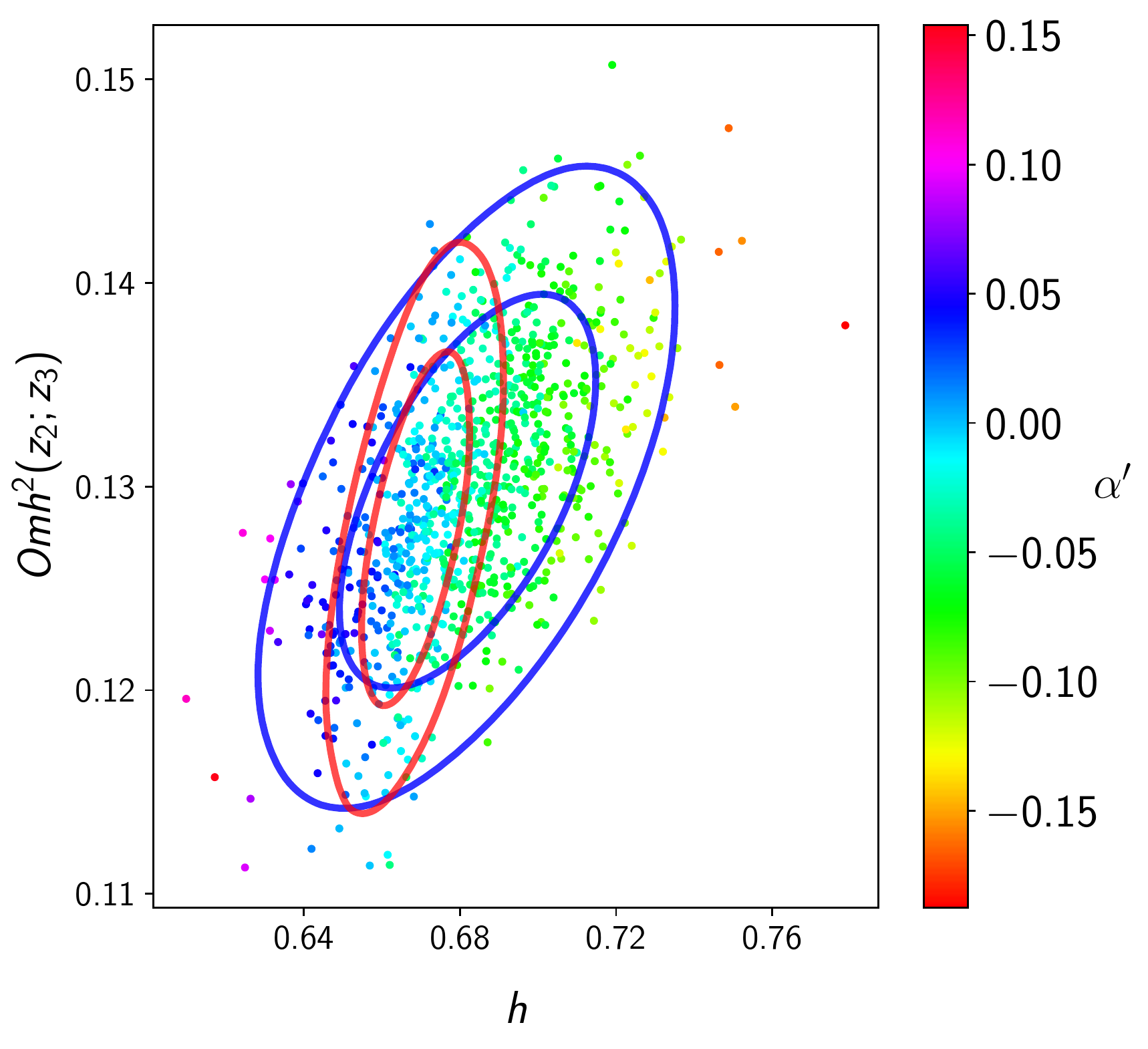}
\caption{Blue lines and 3D scatter color plots described the EMLG model marginalised posterior distributions for EMLG parameter $\alpha'$ in the 
$\{\alpha', Om  h^2(z_i ; z_j),  h_{0}\}$ subspace for $\{z_1,z_2\}$, $\{z_1,z_3\}$ and $\{z_2,z_3\}$. The color code indicates the value of $\alpha'$ labeled by the colour bar. Red lines display 2D marginalised posterior distributions for the $\Lambda$CDM model.}
\label{Omh2}
\end{figure}

In Figure \ref{Omh2} we depict 3D scatter color plots describing the EMLG model marginalised posterior distributions for the EMLG parameter, $\alpha'$, in the $\{\alpha', Om  h^2(z_i ; z_j), h_0\}$ subspace for $\{z_1,z_2\}$, $\{z_1,z_3\}$ and $\{z_2,z_3\}$. In this figure, we see that the 2D marginalised posterior distributions of $\{ Om  h^2(z_i ; z_j), h_0\}$ for the EMLG model (blue contours) are more tilted than the ones for the $\Lambda$CDM model (red contours), implying that a certain increment in $h_0$ would lead to a lesser increment in $Om  h^2(z_i ; z_j)$ in the EMLG model compared to in the $\Lambda$CDM model, and that larger $h_0$ values are allowed for a given $Omh^2$ value provided that $\alpha'$ takes a correspondingly larger negative value, as can be seen from the color gradient indicating $\alpha'$. This implies that the EMLG model compensates for the larger values of $h_0$ by lowering the value of $\alpha'$ and keeps $Om  h^2(z_i ; z_j)$ at lower values. Whereas, in the $\Lambda$CDM model, lowering the value of $Omh^2$ would lead to low $h_0$ values (see Table \ref{tab:priors}) which would exacerbate the tension between the Planck $\Lambda$CDM model and direct $H_0$ measurements. Similarly, increasing the value of $h_0$ would lead to higher $Omh^2$ values but with the difference that a small increment in $h_0$ would lead to relatively larger increments in $Omh^2$ since the red contours for the $\Lambda$CDM model are almost vertical. Indeed, for the $\Lambda$CDM model, in this study we obtain $Omh^2\approx0.128$ along with $h_0\approx0.668$, whereas the recent Planck release gives $Omh^2\approx0.143$ along with $h_0\approx 0.674$. Note that the about 1\% larger value of $h_0$ is accompanied by a roughly 10\% larger value of $Omh^2$.
 
\begin{figure}[!bt]
\captionsetup{justification=raggedright,singlelinecheck=false,font=footnotesize}
\includegraphics[trim = 1mm  1mm 1mm 1mm, clip, width=8.cm, height=5.cm]{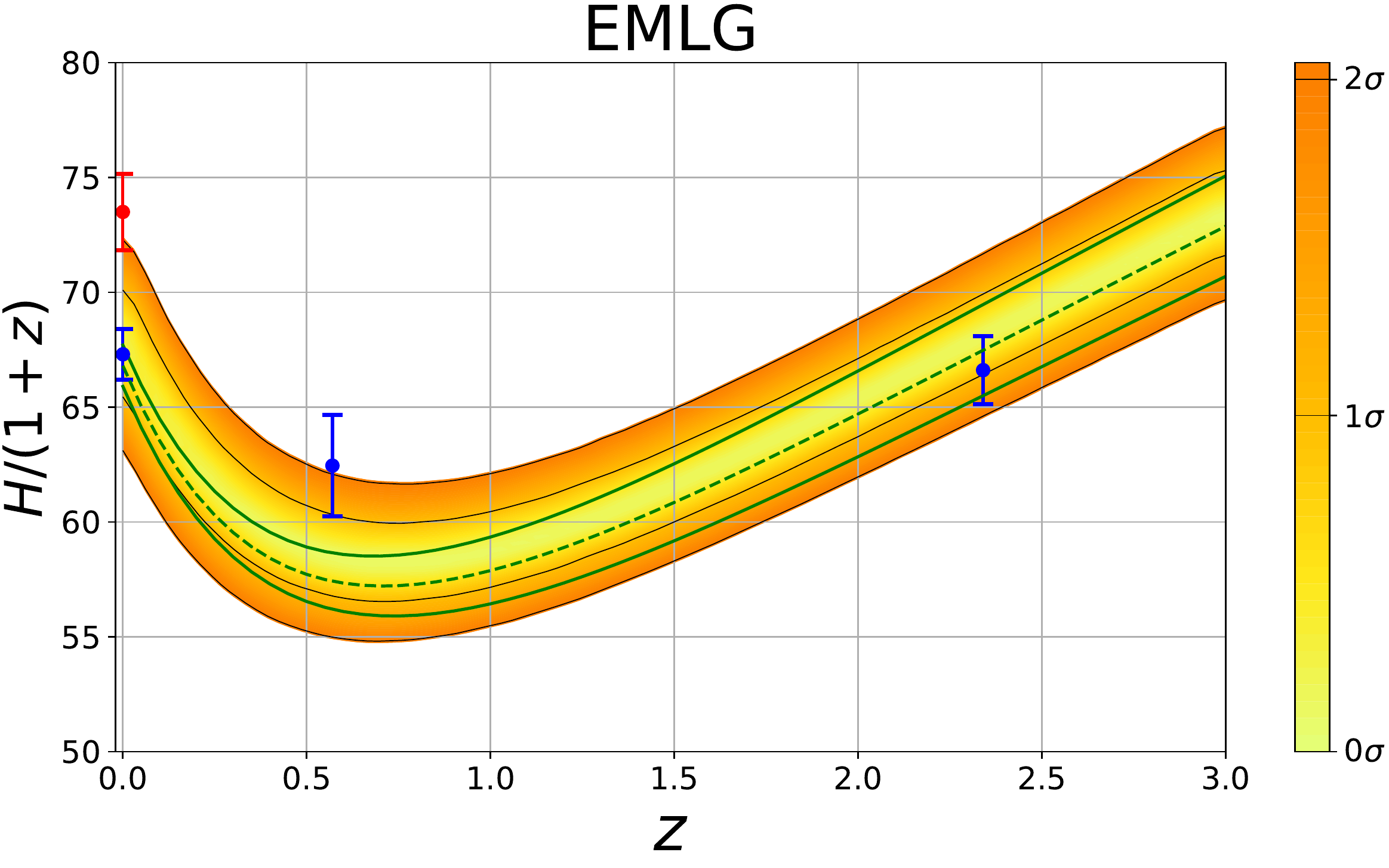}
\includegraphics[trim = 1mm  1mm 1mm 1mm, clip, width=8.cm, height=5.cm]{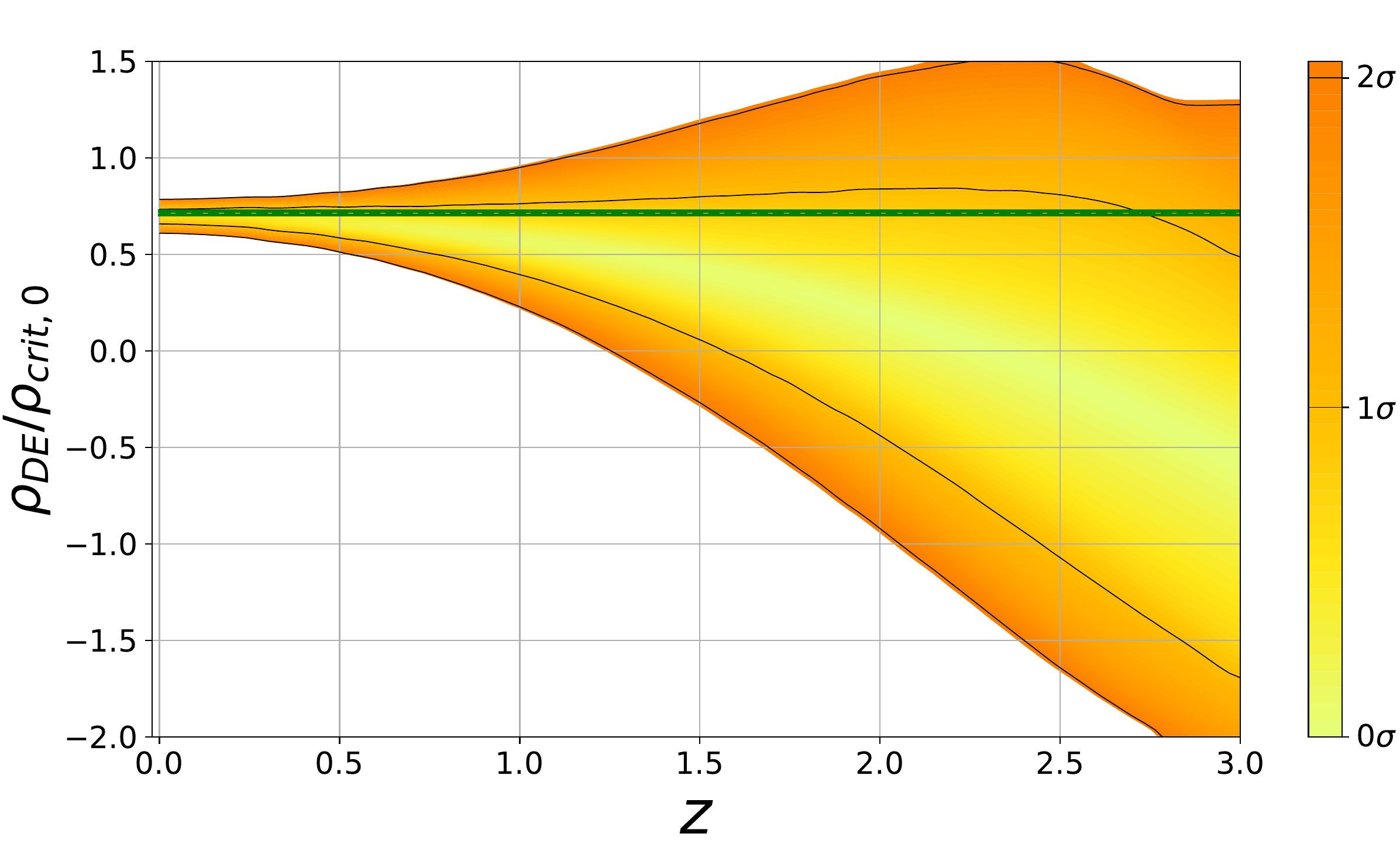}
\caption{\textbf{(Top panel)} $H(z)/(1+z)$ vs. $z$ graph of the EMLG. \textbf{(Bottom panel)} $\rho_{\rm DE}/\rho_{\rm crit,0}$ vs. $z$ graph of the EMLG. For both panels, these show the posterior probability Pr$(g|z)$: the probability of $g$ as normalized in each slice of constant $z$, with color scale in confidence interval values. 
The 1$\sigma$ and 2$\sigma$ confidence intervals are plotted as black lines. Green lines display best-fit values (dotted line) and 1$\sigma$ contour levels
for the $\Lambda$CDM model.}
\label{H_div1+z}
\end{figure}

The data predicts the following constraints on the Hubble constant along with their errors at the 68\% and 95\% confidence levels for the EMLG and the $\Lambda$CDM models:
\begin{align}
H_0= 68.20 \pm 2.13\pm 4.15\,\,  {\rm km}\, {\rm s}^{-1}\,{\rm Mpc}^{-1}, \,\, (\textnormal{EMLG})\\
 H_0= 66.86 \pm 0.90\pm 1.74  \,\,  {\rm km}\, {\rm s}^{-1}\,{\rm Mpc}^{-1}. \,\,  (\Lambda{\rm CDM})
\end{align}
In comparison, the most recent distance-ladder estimates of $H_0$ from the SHOES (SN, $H_0$, for the equation of state of dark energy) project give $H_0= 73.24 \pm 1.74{\rm km\,s}^{-1}{\rm Mpc}^{-1}$ \cite{Riess:2016jrr}, $H_0= 73.48 \pm 1.66{\rm km\,s}^{-1}{\rm Mpc}^{-1}$ \cite{Riess:2018SR} and $H_0 = 73.52 \pm 1.62{\rm km\,s}^{-1}{\rm Mpc}^{-1}$, using Gaia parallaxes \cite{Riess:2018byc}. We note that, at 68\% C.L., $H_0$ values both from the EMLG model and the $\Lambda$CDM model are in tension with these, yet it is worse in the $\Lambda$CDM model. Indeed we see that, at 95\% C.L., the $H_0$ of the EMLG model becomes consistent with these results, while the $H_0$ of the $\Lambda$CDM model remains in tension.

The upper panel of Figure \ref{H_div1+z} displays a subset of the BAO measurements (blue bars) from $z=0$, $z=0.57$ and $z=2.34$ (see \cite{Aubourg:2014yra}) with scalings that illustrate their physical content along with the distance-ladder estimate of $H_0$, the direct observational value (red bar) given in \cite{Riess:2018SR}, and the plot of the posterior probability of $H(z)/(1+z)$, which is the proper velocity between two objects with a constant comoving separation of 1 Mpc, for the EMLG model. We note that the strip (yellow) of $H(z)/(1+z)$ for the EMLG model is consistent with all three BAO data at 1$\sigma$ C.L. (though, marginally with the data from $z=0.57$), whereas it is in tension with the distance-ladder estimate of $H_0$ at 1$\sigma$ but marginally consistent with it at 2$\sigma$ C.L. These indeed are considerable improvement with respect to the $\Lambda$CDM model (green lines displaying the best-fit value (dotted line) and 1$\sigma$ contour levels in the same figure) which is inconsistent with both the BAO data from $z=0.57$ and the distance-ladder estimate of $H_0$ even at 2$\sigma$ C.L. \footnote{Note that in our case $\Lambda$CDM is in tension with the BAO data from $z=0.57$ whereas it is consistent with the one from $z=2.34$ in BOSS \cite{Aubourg:2014yra} and Planck \cite{Aghanim:2018eyx}. The reason being that in our analysis we didn't consider the data from CMB since we omitted radiation in our models.} 

The lower panel of Figure \ref{H_div1+z} shows the probability distribution (yellow tones) of the redshift dependency of the energy density of the effective DE scaled to the critical energy density of the present time Universe, viz., $\rho_{\rm DE}/\rho_{\rm crit,0}$, within 1$\sigma$ and 2$\sigma$ confidence levels for the EMLG model. Whereas the thin green strip in the panel is for the $\Lambda$CDM model at 1$\sigma$ C.L.. We see that the effective DE achieves negative values after few redshifts, namely, we obtain $\rho_{\rm DE}=0$ at $z_*=2.23\pm0.81$ at 1$\sigma$ C.L.. It is noteworthy that this value is in line with that in the BOSS collaboration paper \cite{Aubourg:2014yra} estimating DE with a negative energy density for $z>1.6$ and paper \cite{Sahni:2014dee} suggesting that cosmological models providing effective DE yielding signature change at $z\sim2.4$ to obtain, from the model, $Omh^2$ values consistent with the model-independent estimations.

\section{Conclusions}

We have introduced a new model of Energy-Momentum Squared Gravity, which we call Energy-Momentum Log Gravity (EMLG). It is constructed by the addition of $f(T_{\mu\nu}T^{\mu\nu})=\alpha \ln(\lambda\,T_{\mu\nu}T^{\mu\nu})$, envisaged as correction, to the standard Einstein-Hilbert action with cosmological constant $\Lambda$. We have studied the cosmological solutions of the Friedmann metric that arise from the field equations for this theory of gravitation. Using these solutions we then conducted an investigation into the ways in which the EMLG extension to $\Lambda$CDM addresses the tensions between existing data sets that beset the standard $\Lambda$CDM model. Among the tensions of various degrees of significance reported in the literature, we have focused on the ones discussed in \cite{Aubourg:2014yra,Sahni:2014dee}, which result from the Lyman-$\alpha$ forest measurement of BAO at $z\sim 2.3$ by the BOSS collaboration \cite{Delubac:2014aqe}. It has been argued that this tension can be alleviated in a physically motivated way through a modified gravity theory, rather than as a pure physical DE source within GR \cite{Sahni:2014dee}, since it requires a DE yielding negative energy density values at high redshifts \cite{Aubourg:2014yra,Sahni:2014dee}.

EMLG allows us to find an explicit exact solution for the dust density, $\rho_{\rm m}(z)$,  and thus of $H(z)$ and $\rho_{\rm DE}(z)$ (effective DE), which has allowed us to conduct a detailed theoretical and observational investigation of the model without introducing further simplifications. Following this, upon setting $\Omega_{\rm m,0}=0.28$ and $h_0=0.70$ for both models, we demonstrate analytically that EMLG with $\alpha'=-0.04$ produces effective DE behaving as suggested in \cite{Aubourg:2014yra,Sahni:2014dee} and predicts $Omh^2$ diagnostic values consistent with the model-independent value from observations \cite{Sahni:2014dee}, whereas the value predicted by $\Lambda$CDM exhibits a significant tension with the model-independent value. We have constrained both models against the latest observational data from the combined dataset BAO+SN+CC and then discussed the improvements due to the EMLG modification. It emerges that the data does not rule out the $\Lambda$CDM limit of the model ($\alpha'= 0$), but prefers slightly negative values of the EMLG model parameter ($\alpha'= -0.032\pm 0.043$), which leads to an effective DE indistinguishable from positive $\Lambda$ at low redshifts but results in negative energy density values (i.e., screening of $\Lambda$) for redshift values larger than $z\sim 2.2$, in line with the arguments developed in \cite{Aubourg:2014yra,Sahni:2014dee} for alleviating the tensions relevant to Lyman-$\alpha$ data. We concluded that this feature of the effective DE from the EMLG modification to $\Lambda$CDM arises from the altered redshift dependency of $\rho_{\rm m}$ due to its non-conservation in this model, not from the new type of contributions of it on the right-hand side of the Friedmann equation \eqref{eq:rhoprime-w0}, which yields an effective EoS of a source with constant inertial mass density. We observe further that the EMLG model does this without lowering the values of $\Omega_{\rm m,0}$ and $H_0$ compared to the results from Planck \cite{Planck2015,Aghanim:2018eyx}, and moreover relieves, at some level, the persistent tension with the measurements of $H_0$ within the standard $\Lambda$CDM model. In the case of $\Lambda$CDM, on the other hand, we observed that $Omh^2$ reduces to values consistent with the model independent value, since we did not consider CMB data in our observational analyses, but it happens at the cost of reducing $\Omega_{\rm m,0}$ to values in tension with the Planck result, and also of reducing $H_0$ to values which exacerbate the persistent tension in the measurement of $H_0$.

We see that although our findings are promising in favor of alleviating the tensions considered in this study, they are not yet conclusive. The reason for this is that we have studied only single fluid cosmology, that is we have considered only dust as the material source and excluded the presence of radiation in our model, and equally in $\Lambda$CDM in order to conduct a fair comparison between the models. In order to confirm these initial results, the current study must be extended by the inclusion of radiation together with dust, and then can also be constrained by considering the CMB data along with the other data sets. We have discussed the difficulties of introducing radiation, either by itself or as the second source, in our model and noted a possible way of achieving this, which we reserve for our future works. Finally, we conclude that the current study demonstrates that, through our particular model, EMLG, Energy-Momentum Squared Gravity type extensions to $\Lambda$CDM model are capable of addressing some of the prominent tensions which beset $\Lambda$CDM and merit further investigation.

We would like to close the paper with the following remarks. Our initial motivation for considering $f(T_{\mu\nu}T^{\mu\nu})\propto \ln(\lambda\,T_{\mu\nu}T^{\mu\nu})$ was phenomenological, as gives rise to new contributions by dust on the right-hand side of the Einstein field equations which mimic a source with constant inertial mass density. The corresponding energy density could then change sign at high redshifts as has been suggested for addressing the tension relevant the Lyman-$\alpha$ measurements within the standard $\Lambda $CDM model, although it emerged that our model was able to do so because of the modified redshift dependency of dust due to the non-conservation of energy-momentum tensor. Our model is also expedient as it provides us with an explicit exact solution. On the other hand, one may question the microphysical motivation for such a term; in particular, whether there is a way of realising such a term in the action within a particular field theoretical model that leads to the energy-momentum tensor. For example, naively substituting $T_{\mu\nu}$ with the energy momentum tensor of a scalar field would lead to a quite non-standard (and probably non-analytic) action, which in turn would raise questions about a consistent quantization procedure, the consistency of the corresponding effective field theory, and so on. However, the current paper's primary aim is to highlight the model's cosmological signatures, and in that sense, the work presented here can be understood as a phenomenological contribution to exploring the scope of possibilities. It would be interesting to look for a potential origin of this modification in a theory of fundamental physics and see whether some relationship as between the EMSG of the form $f(T^2)\propto T^2$ \cite{Roshan:2016mbt,Board:2017ign,Akarsu:2018zxl,Nari:2018aqs} and loop quantum gravity \cite{Ashtekar:2006wn,Ashtekar:2011ni} as well as braneworld scenarios \cite{Brax:2003fv}, all of which add quadratic contributions of the matter stresses' energy density to the Friedmann equation, could be found.

\begin{acknowledgements}
\"{O}.A. acknowledges the support by the Turkish Academy of Sciences in the scheme of the Outstanding Young Scientist Award (GEB\.{I}P). 
J.D.B. and C.V.R.B. are supported by the Science and Technology Facilities Council (STFC) of the UK.
J.A.V. acknowledges the support provided by FOSEC SEP-CONACYT Investigaci\'on B\'asica A1-S-21925, and UNAM-DGAPA-PAPIIT  IA102219.
\end{acknowledgements}

\end{document}